%% file: draft_single.tex
\documentclass[12pt,journal,draftcls,a4paper,onecolumn]{IEEEtran} 
\usepackage{dsfont}

\usepackage{subeqnarray}
\usepackage{cleveref}
\usepackage{amssymb}
\usepackage{amsfonts}
\usepackage{amsmath}
\usepackage[dvipdfm]{graphicx}
\usepackage{bmpsize}
\usepackage{amsmath}
\usepackage[all,poly]{xy}
\usepackage{multirow}
\usepackage{algorithm}
\usepackage{algorithmic}
\usepackage{color}
\usepackage[noadjust]{cite}

\title{A Bayesian approach to denoising of single-photon binary images}
\author{Yoann Altmann, Reuben Aspden, Miles Padgett, Steve
McLaughlin \thanks{Yoann Altmann and Steve
McLaughlin are with School of Engineering and Physical Sciences, Heriot-Watt University, Edinburgh,
U.K. (email: \{Y.Altmann;S.McLaughlin\}@hw.ac.uk). Reuben Aspden and Miles Padgett are with School of Physics and Astronomy, University of Glasgow, Glasgow, U.K. (email: \{Reuben.Aspden;Miles.Padgett\}@glasgow.ac.uk)} \thanks{This study was supported by EPSRC via grant EP/J015180/1}}

\newcommand{\bW}{\boldsymbol{W}}

\newcommand{\bPhi}{{\boldsymbol \Phi}}

\input notations_yoann.tex

\begin{document}
\maketitle

\begin{abstract}
This paper discusses new methods for processing images in the photon-limited regime where the number of photons per pixel is binary. We present a new Bayesian denoising method for binary, single-photon images. Each pixel measurement is assumed to follow a Bernoulli distribution whose mean is related by a nonlinear function to the underlying intensity value to be recovered. Adopting a Bayesian approach, we assign the unknown intensity field a smoothness promoting spatial and potentially temporal prior while enforcing the positivity of the intensity. A stochastic simulation method is then used to sample the resulting joint posterior distribution and estimate the unknown intensity, as well as the regularization parameters. We show that this new unsupervised denoising method can also be used to analyze images corrupted by Poisson noise. The proposed algorithm is compared to state-of-the art denoising techniques dedicated to photon-limited images using synthetic and real single-photon measurements. The results presented illustrate the potential benefits of the proposed methodology for photon-limited imaging, in particular with non photon-number resolving detectors.
\end{abstract}

\begin{IEEEkeywords}
Photon-limited imaging, Single-photon detection, image denoising, Bayesian estimation, Markov chain Monte Carlo methods 
\end{IEEEkeywords}

\section{Introduction}
\label{sec:intro}
Single-photon detectors (SPDs) are ubiquitous for applications where the light flux to be analysed is quantified at photonic levels. In particular, SPDs are particularly attractive for imaging applications where the light flux changes rapidly (of the order of picoseconds) or is extremely limited. For instance, the range resolution of SPD-based Lidar systems and their capability to resolve close objects depends on the ability of the detectors to accurately capture the time-of-arrival of photons emitted by fast laser sources \cite{McCarthy2009,Krichel2010,Wallace2010,McCarthy2013,Maccarone2015, Altmann2016a}.

Recent advances in fast SPDs and SPD arrays, coupled with efficient signal/image processing techniques have allowed the development of extreme imaging systems, including first photon \cite{Kirmani2014} and single pixel \cite{Zhang2016,Sun2016}, and ghost \cite{Aspden2013,Aspden2015,Morris2015} imaging systems, among others. 

SPDs can be classified into two groups depending on their ability to quantify a number of detected photons within an elementary time period. Although some detectors are photon-number resolving, in this paper we consider SPDs that can generally only distinguish no detection from at least one detection, such as single-photon avalanche diodes (SPADs), photomultiplier tubes and superconducting nanowire SPDs \cite{Eisaman2011}. Although potentially not too restrictive, such limitations need to be considered when developing/applying statistical methods to analyse data recorded by non photon-number resolving SPDs. Indeed, although the number of photons reaching an SPD within a time period is widely assumed to be Poisson distributed (say of mean $x$), the SPD saturation can have a significant influence on the distribution of the actual photon detections. 

In many imaging applications involving such non photon-revolving SPDs, images are formed by summing binary detection images over several independent realizations and assuming the observed phenomenon is stationary (images identically distributed). By ensuring that the probabilities of detection per acquisition for each pixel (i.e., $1-\exp^{-x}$ assuming the detector has unitary efficiency) are small enough (generally lower than 5\%), the actual distribution of the total number of detected photons in each pixel using $T$ repetitions (e.g. the binomial distribution $\mathcal{B}in\left(T, 1-\exp^{-x}\right)$), can be approximated by a Poisson distribution with mean $T x$. This approximation becomes generally less accurate as $x$ increases (the approximation accuracy depends on $x$ and the number of repetitions $T$ considered). 

If this approach is well adapted to analyze fast low-intensity phenomena for which we naturally have $x<<1$, it requires 1) the intensity field to be constant across the $T$ observations or additional assumptions about its temporal variation (e.g., intensity decay model for fluorescence microscopy \cite{Zwier2004, Jezierska2012_ISBI}) and 2) that $x<<1$ is valid for all the image pixels, which can be problematic when analysing scenes with a high intensity dynamic. Indeed, if the scene includes high intensity regions (i.e., where $x>5\%$), the illumination source has to be reduced (if possible) or attenuation mechanisms (e.g. filters) used to ensure the Poisson noise approximation remains valid across all of the pixels. This automatically and artificially reduces the (already low) probability of detection in the darker regions of the image, potentially unnecessarily, to preserve a tractable observation model. This approach is counterproductive as this decrease is usually compensated for by increasing $T$, the number of repetitions. 
     	  
In this work, and in contrast with most denoising methods developed for photon-limited data, we focus on applications for which the phenomenon can only be observed once and for which we need to infer the intensity field for each individual image. Consequently, the observation model considered assumes the observed images are binary (i.e., either no photon or at least one photon detected). We also consider the case where the detectors are photon-number resolving (i.e., data corrupted by Poisson noise). Here, we focus primarily on binary images even if the proposed methodology can be applied to Poisson data denoising, i.e., for data recorded by photon-number resolving systems. As will be seen in Section \ref{sec:simus_synth}, we show that when using non photon-number resolving systems with relatively high detection probabilities ($x \approx 1, 1-\exp^{-x} \approx 63\%$), it is possible to obtain similar results to photon-number resolving systems. In other words, adopting the appropriate observation model and associated estimation strategy allows for much more efficient data acquisition as it becomes possible to improve the data quality without numerous repetitions (we consider a single detection in this work). However, when saturation is significant, i.e., when $x>>1$, it becomes extremely challenging to accurately estimate the intensity field, in particular using a single frame. In this work, we limit ourselves to $\textrm{E}\left[ x\right]\leq 1$.  


Adopting a classical Bayesian approach, we propose a flexible intensity prior model (see Section \ref{sec:Bayesian_model}) able to capture different sources of intensity fluctuations such as object movement and changes of the illumination conditions. Starting from the observation model of ideal detectors (Poisson likelihood), we present an alternative observation model accounting for detector limitations (Bernoulli likelihood). Both likelihoods are combined with the prior models and a stochastic simulation method (Markov chain Monte Carlo) method is investigated to exploit the resulting posteriors. An important advantage of the proposed method is that it is fully unsupervised and does not require parameter tuning, as the parameters controlling the spatial and temporal regularizations are automatically adjusted during the sampling process. 

The remained of the paper is organized as follow. Section \ref{sec:obs_mod} presents the two observation models considered. The Bayesian models are detailed in Section \ref{sec:Bayesian_model} and the sampling strategy proposed to exploit the resulting posterior distributions is described in Section \ref{sec:MCMC}. Simulation results conducted using synthetic and real single-photon data are
discussed in Sections \ref{sec:simus_synth} and \ref{sec:simus_real}. Conclusions and future work are finally reported in Section \ref{sec:conclusion}.

\section{Observation models}
\label{sec:obs_mod}
 Consider a set of $T$ intensity images $\bfX_t$ of size $N_{\textrm{row}} \times N_{\textrm{row}}$ whose elements $x_{i,j,t}=\left[\bfX_t\right]_{i,j}$ are the unknown average numbers of photons reaching the detector array (composed of $N_{\textrm{row}} \times N_{\textrm{row}}$ detectors regularly spaced) over a given time period. The two observation models are detailed below.

\subsection{Poisson likelihood}
\label{subsec:Poisson}
In the general case where each detector can potentially detect an infinite number of photons (within a given time period $\Delta t$), it is widely acknowledged that the distribution of the photon counts $\pix{i,j}{t}$ in the pixel $(i,j)$ of the $t$th image can be accurately modelled by a Poisson distribution, i.e., 
\begin{eqnarray}
\label{eq:model_poisson}
\pix{i,j}{t}|\left(\eta_{i,j},x_{i,j,t}\right) & \sim & \mathcal{P}\left(\eta_{i,j}x_{i,j,t} \right)
\end{eqnarray} 
where $\eta_{i,j}>0$ is an attenuation factor that stands for the detector sensitivity/efficiency and $\mathcal{P}\left(\eta_{i,j}x_{i,j,t} \right)$ denotes the Poisson distribution with mean $\eta_{i,j}x_{i,j,t}$. In this work we assume that the coefficients $\{\eta_{i,j}\}_{i,j}$ are known (they can usually be estimated during the system calibration) and do not change with time (in the case of video acquisition). Assuming independence between the detectors and the different noise realizations corrupting the images (in particular, we consider non-overlapping acquisition periods) yields the joint likelihood 
\begin{eqnarray}
\label{eq:model_poisson_joint}
f_0(\MATpix|\bfN,\bfX) & = & \prod_{i,j,t}f_0(\pix{i,j}{t}|\eta_{i,j}x_{i,j,t})
\end{eqnarray} 
where $\left[\MATpix\right]_{i,j,t}=\pix{i,j}{t}$, $\left[\bfN\right]_{i,j}=\eta_{i,j}$ and $f_0(\cdot|\eta_{i,j}x_{i,j,t})$ is the Poisson distribution defined in \eqref{eq:model_poisson}.

\subsection{Bernoulli likelihood}
\label{subsec:Bernoulli}
Although the Poisson noise assumption is relevant for many imaging applications SPDs can often only detect at most one photon per pixel within a clock period and need to be reset to potentially detect the next photons reaching the sensor. In the remainder of this paper, we assume that $\Delta t$ corresponds to this clock period (i.e., the smallest temporal sampling period), which also defines the temporal resolution of the imaging system when recording image sequences. In such cases, the detected photon counts, which become binary measurements within a period $\Delta t$, satisfy
\begin{eqnarray}
\pix{i,j}{t} = \left\{
    \begin{array}{ll}
        0 & \mbox{if} ~~ \tilde{y}_{i,j,t} =0 \\
        1 & \mbox{if} ~~ \tilde{y}_{i,j,t}\geq 1
    \end{array}
\right.
\end{eqnarray}
where $\tilde{y}_{i,j,t} \sim \mathcal{P}(\eta_{i,j}x_{i,j,t})$ is the photon count that would be detected by an ideal detector (able to detect an infinite number of photons).
Consequently, in this case the detected photon counts are distributed according to the following Bernoulli distribution
\begin{eqnarray}
\label{eq:lik_bern}
\pix{i,j}{t}| \left(\eta_{i,j}, x_{i,j,t}\right) & \sim & \mathcal{B}er\left(1-\exp^{-\eta_{i,j}x_{i,j,t}} \right)
\end{eqnarray} 
whose mean is given by $1-\exp^{-\eta_{i,j}x_{i,j,t}}$. In a similar fashion to the observation model described in Section \ref{subsec:Poisson}, assuming independence between the detectors  and between noise realizations yields 
\begin{eqnarray}
\label{eq:model_bern_joint}
f_1(\MATpix|\bfN, \bfX) & = & \prod_{i,j,t}f_1(\pix{i,j}{t}|\eta_{i,j}x_{i,j,t}),
\end{eqnarray} 
where $f_1(\cdot|\eta_{i,j}x_{i,j,t})$ denotes the Bernoulli distribution in \eqref{eq:lik_bern}.

%
%

 It is important to mention here that although Poisson and Bernoulli distributions present different shapes and supports, we have 
\begin{eqnarray}
\label{eq:bern_poiss1}
f_1(\pix{i,j}{t}=0| \eta_{i,j}, x_{i,j,t})  & = & f_0(\pix{i,j}{t}=0| \eta_{i,j}, x_{i,j,t}) = \exp^{-\eta_{i,j}x_{i,j,t}}\nonumber
\end{eqnarray}
and 
\begin{eqnarray}
\label{eq:bern_poiss2}
f_1(\pix{i,j}{t}=1| \eta_{i,j}, x_{i,j,t}) & = & \sum_{k=1}^{\infty}f_0(\pix{i}{j}=k| \eta_{i,j}, x_{i,j,t}), 
\end{eqnarray}
which will be useful during the description of the proposed estimation strategy.

\subsection{Faulty sensor and missing data}
In addition to the potential of sensor saturation, we also consider the presence of faulty detectors and missing data within the array. As mentioned in the introduction, the proposed denoising framework exploits the spatial correlation between neighbouring pixels to regularize the denoising problem and the presence of spurious pixels (providing meaningless values) can drastically degrade the algorithm performance. To tackle this problem, we introduce an $N_{\textrm{row}} \times N_{\textrm{row}} \times T$ binary mask $\bfH$ whose entries $h_{i,j,t}$ are $0$ (resp. $1$) when a detector is faulty or data is missing (resp. functioning correctly). 
This mask is assumed to be known and can potentially vary with time. When considering the presence of faulty detectors/missing data, the likelihood functions \eqref{eq:model_poisson_joint} and \eqref{eq:model_bern_joint} become
\begin{eqnarray}
\label{eq:model_poisson_final_joint}
\tilde{f}_0(\MATpix|\bfN,\bfX, \bfH) &= &\prod_{i,j,t} (1-h_{i,j,t})\delta(\pix{i,j}{t}) + h_{i,j,t}f_0(\pix{i,j}{t}| \eta_{i,j}, x_{i,j,t})
\end{eqnarray}
and
\begin{eqnarray}
\label{eq:model_bern_final_joint}
\tilde{f}_1(\MATpix|\bfN,\bfX, \bfH) &= &\prod_{i,j,t} (1-h_{i,j,t})\delta(\pix{i,j}{t}) + h_{i,j,t}f_1(\pix{i,j}{t}| \eta_{i,j}, x_{i,j,t})
\end{eqnarray} 
respectively, where $\delta(\cdot)$ denotes the Dirac delta function and where the observations of faulty pixels (whose positions are known) are arbitrarily set to $0$ before applying the proposed method. These values are not used during the denoising process. 

The proposed methodology does not assume a particular structure for $\bfH$. Although the observation models in \eqref{eq:model_poisson_final_joint} and \eqref{eq:model_bern_final_joint} could be used for restoration of sparsely and/or irregularly sampled images, we restrain ourselves to cases where the number of zero entries in $\bfH$ is small compared to the number of pixels/detectors. The very interesting and more challenging problem of sparsely sampled images constructed from sparse single-photon data outwith the scope of this paper.

The next section describes the Bayesian model and associated estimation strategy proposed to solve the denoising problem considered here; that is, the estimation of the unknown and non-stationary intensity field $\bfX$ from the observed set of photon counts in $\MATpix$.  

\section{Bayesian model}
\label{sec:Bayesian_model}
\subsection{Intensity field modelling}
As in most ill-posed inverse problems which need to be regularized, the choice of the regularization or prior model considered for image restoration is crucial both in terms of quality of image recovery and the resulting computational complexity. In this work we investigate a Bayesian model coupled with an efficient simulation method which allows the estimation of denoised images but that can also provide information about the denoising uncertainty via measures of uncertainty from the posterior distribution of interest. Consequently, we investigate an intensity prior model which allows the use of an simple simulation strategy to exploit the posterior distribution. As will be shown in Sections \ref{sec:simus_synth} and \ref{sec:simus_real}, the prior models presented in this section not only facilitate the estimation strategy but are also flexible enough to compete with standard regularizations used to denoise images corrupted by Poisson noise (e.g., total-variation \cite{Figueiredo2010,Harmany2012} or Gaussian MRFs \cite{Orieux2012}).
 
It is well known that gamma distributions are conjugate priors for the means of Poisson distributions, which makes them particularly attractive to denoise images corrupted by Poisson noise. Moreover, as will be further discussed in Section \ref{sec:MCMC}, gamma distributions remain conjugate priors when considering saturating sensors (i.e., assuming \eqref{eq:lik_bern}), which is particularly convenient in simplifying the denoising problem when the data are Bernoulli distributed. 

\begin{figure}[ht]
\centering
\includegraphics[width=\columnwidth]{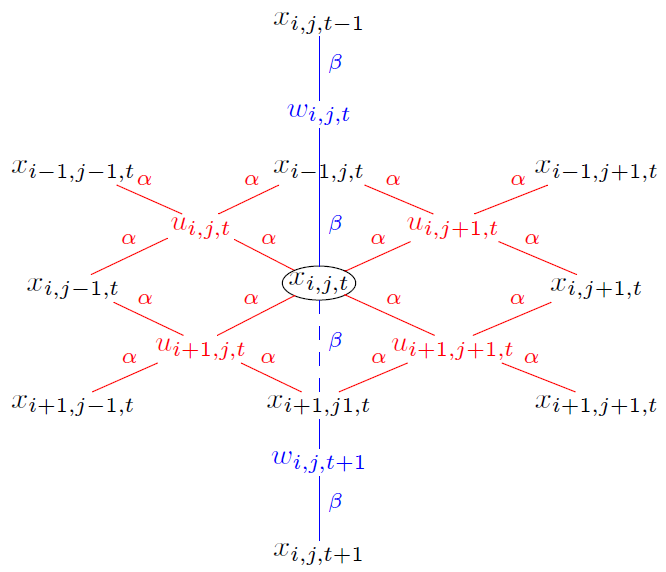}
\caption{Proposed $1$st order GMRF neighborhood structure $\forall (i,j,t) \in \mathcal{V}_{\bfX}\times \mathcal{T}$. The red (resp. blue) sub-graph highlights the spatial (resp. temporal) neighborhood structure. For the pixels at the boundaries of the image/ image sequence, we assume the images to be cyclic spatially (e.g., $x_{0,j,t}=x_{N_{\textrm{row}},j,t}$) and set $x_{i,j,0}=x_{i,j,T+1}=\gamma, \forall (i,j) \in \mathcal{V}_{\bfX}$. The temporal boundary condition $\gamma$ is set arbitrarily to the empirical mean of the observed images but the image sequence can also be assumed to be cyclic.}
\label{fig:neighbour_GMRF2}
\end{figure}

Due to the spatial organization of images, we expect the values of $x_{i,j,t}$ to vary smoothly from one pixel to another. Moreover, if an image sequence is considered, it might be relevant to capture the temporal correlation between successive images to improve the denoising performance, especially since the sampling period can be extremely short (of the order of nanoseconds or less). In order to model this behaviour, we consider an extended prior model such that the resulting prior for $\bfX$ is a hidden gamma-MRF (GMRF) \cite{Dikmen2010}. 
In a similar fashion to \cite{Altmann2016a}, we introduce $T$ auxiliary matrices $\bfU_t$ of size $(N_{\textrm{row}}+1) \times (N_{\textrm{col}}+1)$  with elements $u_{i,j,t} \in \mathbb{R}^+$ and $T+1$ additional auxiliary images $\bfW_t$ of size $N_{\textrm{row}} \times N_{\textrm{row}}$. We then define a tripartite conditional independence graph between $\bfX$, $\bfU=\{\bfU_t\}$ and $\bfW=\{\bfW_t\}$ such that each $x_{i,j,t}$ of $\bfX_t$ is connected to four (spatial) neighbors of $\bfU_t$ and two temporal neighbors in $\bfW_t$ and $\bfW_{t+1}$. This $1$st order neighbourhood structure is depicted in Fig. \ref{fig:neighbour_GMRF2}, where we notice that any given $x_{i,j,t}$ and $x_{i+1,j,t}$ are $2$nd order neighbors via $u_{i+1,j,t}$ and $u_{i+1,j+1,t}$. 
Similarly, $x_{i,j,t}$ and $x_{i,j,t+1}$ are $2$nd order neighbors via $w_{i,j,t+1}$. Following the general GMRF model proposed in \cite{Dikmen2010} and specified here by the neighbourhood structure depicted in Fig. \ref{fig:neighbour_GMRF2}, we assign $(\bfX,\bfU,\bfW)$ a (constrained) GMRF prior, and obtain the joint prior $f(\bfX,\bfU,\bfW|\alpha,\beta)$ (see \cite{Dikmen2010} for the GMRF formulation adopted here). This prior model explicitly depends on the value of the hyperparameters $\alpha>0$ and $\beta>0$, which here act as regularization parameters that control the amount of spatial ($\alpha$) and temporal ($\beta$) smoothness enforced by the GMRF. For brevity, we assume that these parameters are fixed and constant across the image sequence in the remainder of this Section. 
However, following an empirical Bayesian approach, in the results presented in Sections \ref{sec:simus_synth} and \ref{sec:simus_real}, the value of $(\alpha,\beta)$ is adjusted automatically (either for each image or for all the images) during the early iterations of the sampler by maximum marginal likelihood estimation (the interested reader is invited to consult \cite{Pereyra2014ssp,Altmann2016a} for details about the estimation of $\alpha$ (or $(\alpha,\beta)$)).

Exploiting the proposed neighbourhood structure yields  
\begin{subeqnarray}
\label{eq:prior2_r}
\slabel{eq:prior2_r2}
x_{i,j,t}|\bfU_t,\bfW,\alpha,\beta &\sim & \mathcal{G}_{\mathcal{X}}\left(\alpha+\beta, \tilde{x}_{i,j,t}\right)  \\
\slabel{eq:prior2_r3}
u_{i,j,t}|\bfX_t,\alpha &\sim & \mathcal{IG}\left(\alpha,\alpha \tilde{u}_{i,j,t}\right)\\
\slabel{eq:prior2_r4}
w_{i,j,t}|\bfX,\beta &\sim & \mathcal{IG}\left(\beta,\beta \tilde{w}_{i,j,t}\right)
\end{subeqnarray}
where
\begin{eqnarray*}
\tilde{x}_{i,j,t} & = & 4/\alpha\left(u_{i,j,t}^{-1} + u_{i-1,j,t}^{-1} + u_{i,j-1,t}^{-1} + u_{i-1,j-1,t}^{-1}\right)^{-1} + 2/\beta\left(w_{i,j,t}^{-1} + w_{i,j,t+1}^{-1}\right)^{-1}\nonumber\\
\tilde{u}_{i,j,t} & = & \left(x_{i,j,t} + x_{i+1,j,t} + x_{i,j+1,t} + x_{i+1,j+1,t}\right)/4\\
\tilde{w}_{i,j,t} & = &\left(x_{i,j,t-1} + x_{i,j,t}\right)/2,
\end{eqnarray*}
and $\mathcal{G}_{\mathcal{X}}\left(\cdot,\cdot\right)$ denotes a gamma distribution restricted to $\mathcal{X}$ (this distribution reduces to a standard gamma distribution with $\mathcal{X}=(0,+\infty)$) and $\mathcal{IG}\left(\cdot,\cdot\right)$ denotes an inverse-gamma distribution. 
If a single image or independent images are considered, the GMRF-based prior $f(\bfX,\bfU,\bfW|\alpha,\beta)$ can be simplified by removing the auxiliary variables $\bfW$ and by considering the prior model $f(\bfX,\bfU|\alpha)$, as in \cite{Altmann2016a}. In that case, the neighborhood structure reduces to the red subgroup of Fig. \ref{fig:neighbour_GMRF2} and we obtain
\begin{subeqnarray}
\label{eq:prior1_r}
\slabel{eq:prior1_r2}
x_{i,j,t}|\bfU_t,\alpha &\sim & \mathcal{G}_{\mathcal{X}}\left(\alpha, \dfrac{\bar{x}_{i,j,t}}{\alpha}\right) \\
\slabel{eq:prior1_r3}
u_{i,j,t}|\bfX_t,\alpha &\sim & \mathcal{IG}\left(\alpha,\alpha \tilde{u}_{i,j,t}\right)
\end{subeqnarray}
where
\begin{eqnarray*}
\bar{x}_{i,j,t} & = & 4\left(u_{i,j,t}^{-1} + u_{i-1,j,t}^{-1} + u_{i,j-1,t}^{-1} + u_{i-1,j-1,t}^{-1}\right)^{-1}
\end{eqnarray*}

In addition to their flexibility, one of the main motivations for considering GMRFs here is the fact that they make the sampling strategy cosier and thus the inference process (using the conjugacy of \eqref{eq:prior2_r2} and \eqref{eq:model_poisson_final_joint} or \eqref{eq:prior1_r2} and \eqref{eq:model_bern_final_joint} (see Eq. \eqref{eq:post_X_1})), while introducing spatial and temporal dependencies between the neighbouring intensities.

\subsection{Joint posterior distributions}
Now that we have defined the prior model for the unknown image or images to be recovered, we can derive the posterior distribution of $(\bfX,\bfU)$ or $(\bfX,\bfU,\bfW)$ (depending on whether temporal correlation is considered), given the observations $\MATpix$, and the fixed model parameters/hyperparameters $\bPhi=\{\bfH,\bfN,\alpha,\beta\}$ and the observation model considered. Using Bayes' rule, we obtain
\begin{eqnarray}
\label{eq:post_indep}
f_m(\bfX,\bfU|\MATpix,\bPhi) \propto \tilde{f}_m(\MATpix|\bfN,\bfX, \bfH)f(\bfX,\bfU|\alpha)
\end{eqnarray}
with $m=0$ (Poisson noise) or $m=1$ (Bernoulli realizations) for a single or independent images and
\begin{eqnarray}
\label{eq:post_correl}
f_m(\bfX,\bfU,\bfW|\MATpix,\bPhi) \propto \tilde{f}_m(\MATpix|\bfN,\bfX, \bfH)f(\bfX,\bfU,\bW|\alpha,\beta)
\end{eqnarray}
when considering temporally correlated images, where $\tilde{f}_m(\MATpix|\bfN,\bfX, \bfH)$ is given either by \eqref{eq:model_poisson_final_joint} (ideal detector) or \eqref{eq:model_bern_final_joint} (saturated detector). 
The next paragraph details the Markov chain Monte Carlo (MCMC) method proposed to sample the posteriors of interest \eqref{eq:post_indep} and \eqref{eq:post_correl} and subsequently estimate the unknown intensity field $\bfX$.

\section{Estimation strategy}
\label{sec:MCMC}
In this work we adopt a simulation based strategy to approximate, for each model, the marginal posterior mean or minimum mean square error (MMSE) estimator of $\bfX$, i.e., 
\begin{eqnarray}
\label{eq:MMSE}
\widehat{\bfX} = \textrm{E}\left[\bfX|\MATpix, \bPhi\right], 
\end{eqnarray}
where the auxiliary variables $\bfU$ (and $\bfW$ when considering correlated frames) have been marginalized. Note that by considering the marginal posterior $f_m(\bfX|\MATpix, \bPhi)$ the corresponding measures of uncertainty automatically accounts for the uncertainty induced by the unknown auxiliary variables in $\bfU$ (and $\bfW$). 

Although marginalizing analytically $\bfU$ and $\bfW$ from \eqref{eq:post_indep} and \eqref{eq:post_correl} is possible using the structure of the GMRFs $f(\bfX,\bfU|\alpha)$ or $f(\bfX,\bfU,\bfW|\alpha,\beta)$, estimating $\bfX$ directly from $f_m(\bfX|\MATpix, \bPhi)$ is challenging due to the complexity of this non-standard and high-dimensional distribution. Fortunately, for the two observation models and the two prior models considered, \eqref{eq:MMSE} can be efficiently approximated with arbitrarily large accuracy by Monte Carlo integration. More precisely, it is possible to compute \eqref{eq:MMSE} by first using an MCMC computational method to generate samples
asymptotically distributed according to \eqref{eq:post_indep} or \eqref{eq:post_correl}, and subsequently using these samples to approximate the required marginal expectation. 

Here we propose a Gibbs/Metropolis-within-Gibbs sampler to simulate samples from the full posterior of interest, as this type of MCMC method is particularly suitable
for models involving hidden Markov random fields \cite[Chap. 10]{Robert2004}. The output of this algorithm is a Markov chain of $N_{\textrm{MC}}$ samples $\bfX^{(1)},\ldots,\bfX^{(N_{\textrm{MC}})}$
that are asymptotically distributed according to the marginal posterior
distribution $f_m(\bfX|\MATpix, \bPhi)$. The first $N_{\textrm{bi}}$ samples of
these chains correspond to the so-called burn-in transient
period and should be discarded (the length of this period can
be assessed visually from the chain plots or by computing
convergence tests \cite{Robertmcmc}). The remaining $N_{\textrm{MC}} - N_{\textrm{bi}}$ samples are used to approximate the Bayesian estimator \eqref{eq:MMSE} as follows
\begin{eqnarray}
\widehat{\bfX} = \dfrac{1}{N_{\textrm{MC}} - N_{\textrm{bi}}}\sum_{t=N_{\textrm{bi}} + 1}^{N_{\textrm{MC}}}\bfX^{(t)}. 
\end{eqnarray}
The remainder of this section details the main steps of the proposed samplers, depending on the observation model considered. The main steps of the proposed PID-GMRF and BID-GMRF (for Poisson and Bernoulli image denoising using GMRF) are summarized in Algo. \ref{algo:algo1} below.

\subsection{Sampling the auxiliary variables}
Since the auxiliary variables $\bfU$ and $\bfW$ do not appear in the likelihoods \eqref{eq:model_poisson_final_joint} and \eqref{eq:model_bern_final_joint}, sampling from their conditional distributions reduces to sampling from \eqref{eq:prior1_r3} (single or independent images) or \eqref{eq:prior2_r3}-\eqref{eq:prior2_r4} (correlated images), whatever the observation model considered (e.g., Poisson or Bernoulli model). Thanks to the structure of the GMRFs considered, the elements of $\bfU$ and $(\bfU,\bfW)$ are a posteriori mutually independent (conditioned on the value of $\bfX$) and can thus be updated in a parallel manner. 

\clearpage 
\begin{algogo}{Poisson/Bernoulli image denoising (PID-GMRF/BID-GMRF)}
\label{algo:algo1}
 \begin{algorithmic}[1]
			\STATE \underline{Fixed input parameters:} $\bfH,\alpha,\beta$, number of burn-in iterations $N_{\textrm{bi}}$, total number of iterations $N_{\textrm{MC}}$, temporal correlation binary label $z_{3D} \in (0,1)$, observation model $m$ ($0$ for Poisson and $1$ for Bernoulli).
			\STATE \underline{Initialization ($k=0$)}
			\STATE Set $\bfX^{(0)},\bfU^{(0)}$ and $\bfW^{(0)}$ 
			\STATE \underline{Iterations ($1 \leq k \leq N_{\textrm{MC}}$)}
			\STATE Sample $\bfU^{(k)} \sim f(\bfU^{(k)}|\bfX^{(k-1)},\alpha)$ in \eqref{eq:prior2_r3}
			\IF{$z_{3D}=1$}
				\STATE Sample $\bfW^{(k)} \sim f(\bfW|\bfX^{(k-1)},\beta)$ in \eqref{eq:prior2_r4}
			\ENDIF
			\FOR{$(i,j,t) \in {V}_{\bfX} \times \mathcal{T}$}
				\IF{$h_{i,j,t}=0$ (faulty pixel)}
					\STATE Sample $x_{i,j,t}^{(k)}$ using $(\bfU^{(k)},\bfW^{(k)})$ and \eqref{eq:prior1_r2} ($z_{3D}=0$) or \eqref{eq:prior2_r2} ($z_{3D}=1$)
				\ELSIF{m=0}
					\STATE Sample $x_{i,j,t}^{k}$ using $(\bfU^{(k)},\bfW^{(k)})$ and \eqref{eq:post_X_1} ($z_{3D}=0$) or \eqref{eq:post_X_2} ($z_{3D}=1$)
				\ELSE 
					\STATE Sample $x^*$ using \eqref{eq:post_X_1} ($z_{3D}=0$) or \eqref{eq:post_X_2} ($z_{3D}=1$)
					\STATE Compute $\rho$ and $\mu$ using \eqref{eq:accept_2D} ($z_{3D}=0$) or \eqref{eq:accept_3D} ($z_{3D}=1$)
					\STATE Sample $\nu \sim \mathcal{U}_{(0,1)}(\nu)$
					\IF{$\nu<\mu$}
						\STATE Set $x_{i,j,t}^{(k)}=x^*$
					\ELSE
						\STATE Set $x_{i,j,t}^{(k)}=x_{i,j,t}^{(k-1)}$
					\ENDIF
				\ENDIF
			\ENDFOR
		\STATE Optional: Update $\alpha$ or $(\alpha,\beta)$ using \cite{Pereyra2014ssp}.
		\STATE Set $k = k+1$.
\end{algorithmic}
\end{algogo}

\subsection{Sampling $\bfX$}
Similarly, it is easy to show that for a given realization of $\bfU$ (and $\bfW$), the elements of $\bfX$ are a posteriori independent and can be updated simultaneously. If a given pixel $(i,j,t)$ is faulty or does not contain meaningful data, i.e., $h_{i,j,t}=0$, its corresponding unknown intensity value does not appear in \eqref{eq:model_poisson_final_joint} nor in \eqref{eq:model_bern_final_joint}. Consequently, sampling such intensity values reduces to sampling from \eqref{eq:prior1_r2} (single or independent images) or \eqref{eq:prior2_r2} (correlated images). Sampling from truncated gamma distributions can be done efficiently via rejection sampling by sampling from non-truncated gamma distributions, in particular when the non-truncated distribution is mainly concentrated in $\mathcal{X}$.

Consider a valid pixel $(i,j,t)$ following the observation model \eqref{eq:model_poisson}. It is easy to obtain using the Poisson-gamma conjugacy that
\begin{eqnarray}
\label{eq:post_X_1}
x_{i,j,t}|y_{i,j,t},\bfU_t, \bPhi \sim \mathcal{G}_{\mathcal{X}}\left(\alpha+y_{i,j,t}, \dfrac{\bar{x}_{i,j,t}}{1+ \bar{x}_{i,j,t}\eta_{i,j}}\right) 
\end{eqnarray}
for a single image or independent images and
\begin{eqnarray}
\label{eq:post_X_2}
x_{i,j,t}|y_{i,j,t},\bfU_t, \bfW,\bPhi \sim \mathcal{G}_{\mathcal{X}}\left(\alpha+\beta+y_{i,j,t}, \dfrac{\tilde{x}_{i,j,t}}{1 + \tilde{x}_{i,j,t}\eta_{i,j}}\right) 
\end{eqnarray}
for correlated images. These distributions, denoted as $f_0^{2D}(x_{i,j,t}|y_{i,j,t},\bfU_t, \bPhi)$ and\\
$f_0^{3D}(x_{i,j,t}|y_{i,j,t},\bfU_t, \bfW,\bPhi)$, respectively, can also be sampled from via rejection sampling. 

Consider now a valid pixel $(i,j,t)$ following the observation model \eqref{eq:lik_bern}. We are interested in the expression of $f_1^{2D}(x_{i,j,t}|y_{i,j,t},\bfU_t, \bPhi)$ and $f_1^{3D}(x_{i,j,t}|y_{i,j,t},\bfU_t, \bfW,\bPhi)$, the conditional distributions of $x_{i,j,t}$ using the 2D and 3D GMRFs respectively. By recalling that \eqref{eq:bern_poiss1} and \eqref{eq:bern_poiss2}, it can be shown that 
\begin{eqnarray}
f_1^{2D}(x_{i,j,t}|y_{i,j,t}=0,\bfU_t, \bPhi)  = f_0^{2D}(x_{i,j,t}|y_{i,j,t}=0,\bfU_t, \bPhi)\nonumber
\end{eqnarray}
and
\begin{eqnarray}
f_1^{3D}(x_{i,j,t}|y_{i,j,t}=0,\bfU_t, \bfW,\bPhi) =  f_0^{3D}(x_{i,j,t}|y_{i,j,t}=0,\bfU_t, \bfW,\bPhi)
\end{eqnarray}
are truncated gamma distributions and that $f_1^{2D}(x_{i,j,t}|y_{i,j,t}=1,\bfU_t, \bPhi)$ and $f_1^{3D}(x_{i,j,t}|y_{i,j,t}=1,\bfU_t, \bfW,\bPhi)$ are infinite mixtures of gamma distributions which are less trivial to sample from. To tackle this problem, we introduce a Metropolis-Hastings move to update $x_{i,j,t}$ under a Bernoulli observation assumption. More precisely, for a given valid pixel $(i,j,t)$ at the $k$th iteration of the sampler, we can use a so-called proposal distribution $q(\cdot)$ defined on $\mathcal{X}$ to generate a candidate $x^*$. This candidate is then accepted with probability $\mu=\textrm{min}(\rho,1)$ where 
\begin{eqnarray}
\label{eq:accept_2D}
\rho = \dfrac{f_1^{2D}(x^*|y_{i,j,t},\bfU_t, \bPhi)q(x_{i,j,j}^{(k)})}{f_1^{2D}(x_{i,j,j}^{(k)}|y_{i,j,t},\bfU_t, \bPhi)q(x^*)}
\end{eqnarray}
and 
\begin{eqnarray}
\label{eq:accept_3D}
\rho = \dfrac{f_m^{3D}(x^*|y_{i,j,t},\bfU_t, \bfW, \bPhi)q(x_{i,j,j}^{(k)})}{f_m^{3D}(x_{i,j,j}^{(k)}|y_{i,j,t},\bfU_t, \bfW, \bPhi)q(x^*)}
\end{eqnarray}
using the 2D and 3D GMRFs, respectively. Otherwise, we set $x_{i,j,t}^{(k)}=x_{i,j,t}^{(k-1)}$.
Here, to avoid additional algorithmic complexity (e.g., tuning the variances of Gaussian random walks), we use as proposal distributions the conditional distributions obtained under Poisson noise assumption \eqref{eq:post_X_1} or \eqref{eq:post_X_2}, depending on the scenario considered (independent/correlated images). Using this choice of proposal, 1) when $y_{i,j,t}=0$, we obtain $\rho=1$ and the Metropolis-Hastings step reduces to a Gibbs step and 2) in practice we have observed that this choice leads to satisfactory acceptance rates ($\rho>0.6$ for the pixels such such that $y_{i,j,t}=1$) for all the results presented in Sections \ref{sec:simus_synth} and \ref{sec:simus_real}.

In this Section, we considered two intensity prior models, depending on whether the $T>1$ observed images are assumed to be temporally correlated or not. As the underlying intensity field is the same if the detectors saturate or not, a single prior model is considered when analysing Poisson or Bernoulli images. We then detailed a single sampling strategy to exploit the posterior distributions of the different scenarios. When the data are Bernoulli observations, accept/reject procedures are only required for the pixels where a detection occurs, i.e., $y_{i,j,t}=1$, which, in the case of low illumination images ($\textrm{E}\left[y_{i,j,t} \right]<<1$), represent a small fraction of the pixels. In the case of Poisson data, the proposed Metropolis-within-Gibbs sampler reduces to a standard, yet highly parallelizable, Gibbs sampler. The following Sections illustrate the potential benefits of the proposed method through a series of experiments conducted using synthetic and real single-photon images.

\section{Simulations using synthetic images}
\label{sec:simus_synth}
In this Section we investigate the performance of the proposed methods and the effect of the detector saturation on the intensity estimation performance through simulations conducted with synthetic images. First, we compare the performance of the proposed algorithms with existing methods when denoising a single image. Then we assess the benefits of the proposed 3D GMRF, when denoising a sequence of images.

\subsection{Single image denoising}

\begin{figure}[h!]
\begin{minipage}[b]{.49\linewidth}
  \centering
\includegraphics[width=4.2cm]{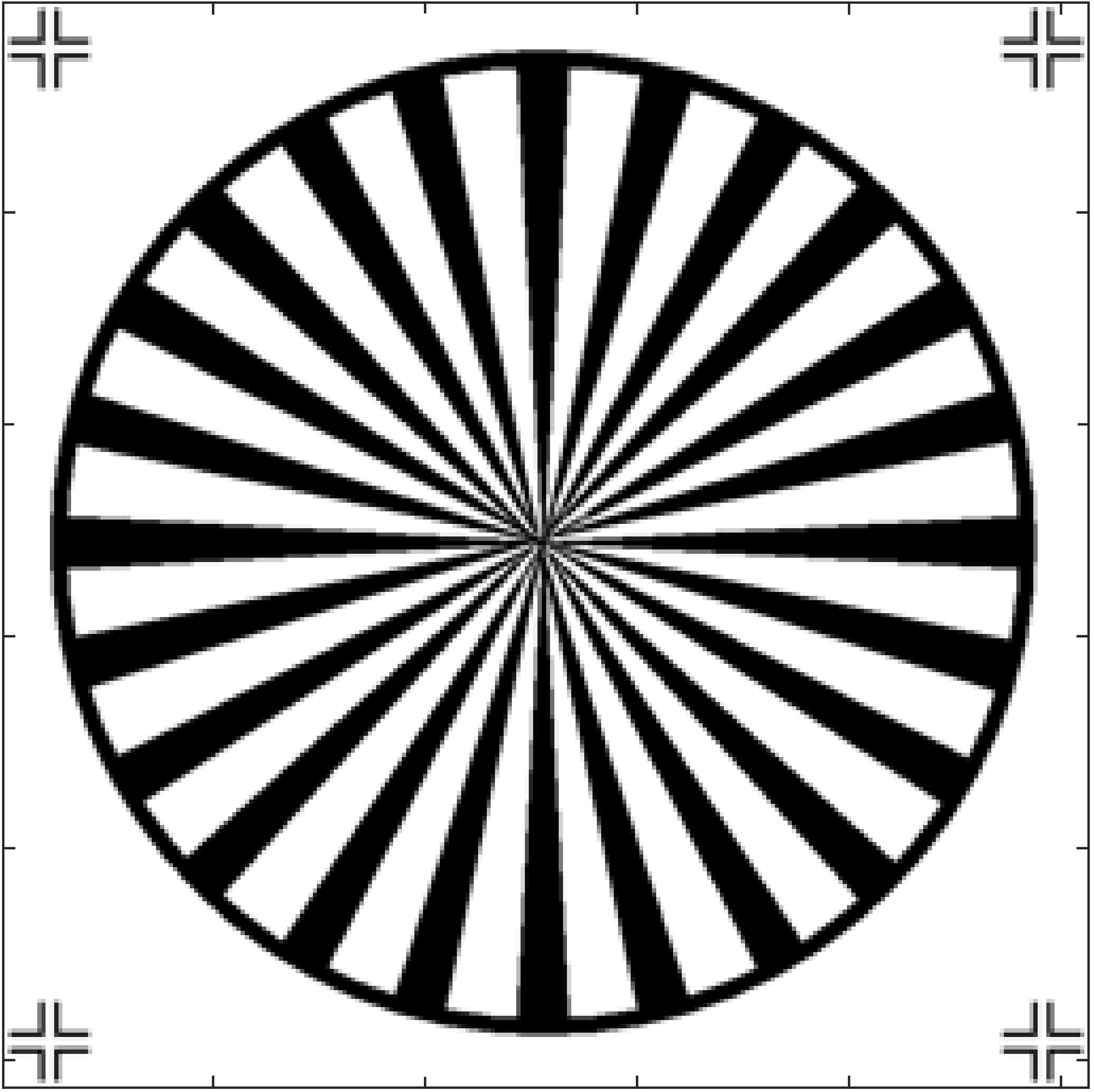}
 \vspace{-0.2cm}
  \centerline{(a) }\medskip
\end{minipage}
\hfill
\begin{minipage}[b]{0.49\linewidth}
  \centering
\includegraphics[width=4.2cm]{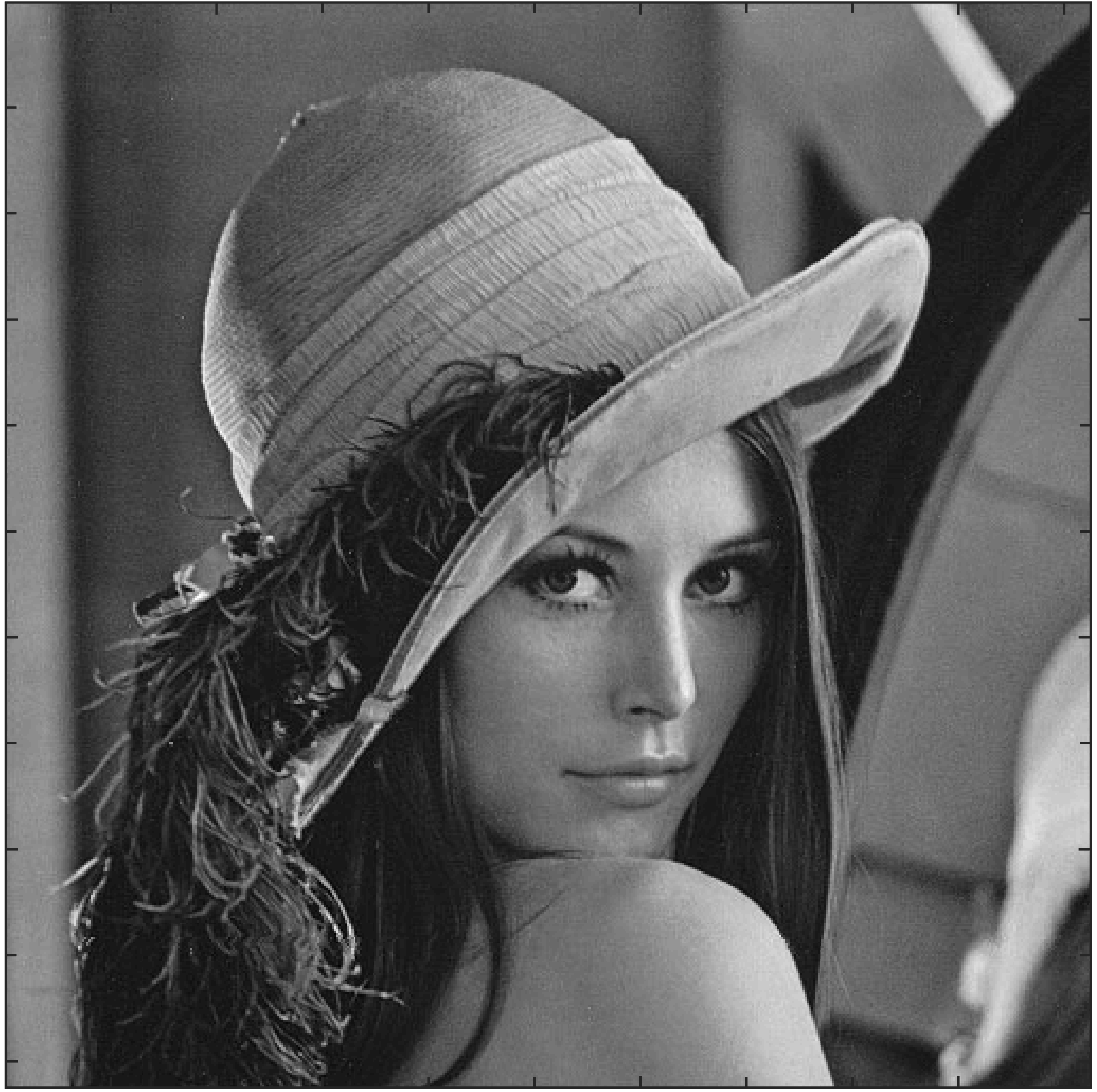}
 \vspace{-0.2cm}
  \centerline{(b)}\medskip
\end{minipage}
\vspace{-0.6cm} \caption{(a): First image $I_1$ composed of a piece-wise constant intensity profile. (b) Second image $I_2$ considered, which presents smoother intensity variations.} \label{fig:single_im}
\end{figure}

We evaluate the proposed methods in denoising the two test images depicted in Fig. \ref{fig:single_im}. The first image of size $256 \times 256$ (circular pattern) and denoted $I_1$, presents a piece-wise constant intensity profile while the second image $I_2$, of size $512 \times 512$ presents smoother intensity variations. In all the experiments presented in this section, for fair comparisons to methods which cannot handle missing data, we assume that all pixels are observed. We then repeated the same experiments with up to $0.1\%$ of missing data/outliers and did not observe noticeable changes in the denoising performance of the proposed methods. Here, we  used $\eta_{i,j}=1, \forall (i,j)$. The two original images have been scaled such that the expected number of counts (averaged over the image pixels) $\textrm{E}\left[x_{i,j}\right] \in \{2.5\%;5\%;10\%;50\%;80\%;100\%\}$. For each value of $\textrm{E}\left[x_{i,j}\right]$, $T=20$ independent noisy images have been generated using the model described by \eqref{eq:lik_bern}. To compare the results with those obtained when the data are corrupted by Poisson noise, we also generate data using \eqref{eq:model_poisson}. Table \ref{tab:data_analysis} gathers details about the mean observed intensity values (averaged over all the pixels and the $T=20$ noise realizations) for the different scenarios. In contrast to the corruption by Poisson noise, $\textrm{E}\left[y_{i,j}\right]$ is much smaller than $\textrm{E}\left[x_{i,j}\right]$ for large values of $\textrm{E}\left[x_{i,j}\right]$ when considering Bernoulli noise. However, this difference (which also depends on the distribution of $x_{i,j}$ across the pixels) reduces for small values of $\textrm{E}\left[x_{i,j}\right]$. As expected, for $\textrm{E}\left[x_{i,j}\right]< 5\%$, the Bernoulli and Poisson distribution are very similar.  

\begin{table}[h!]
\renewcommand{\arraystretch}{1.2}
\begin{footnotesize}
\begin{center}
\begin{tabular}{|c|c|c|c|c|c|c|c|}
\cline{3-8}
\multicolumn{2}{c|}{} &  \multicolumn{6}{|c|}{$\textrm{E}\left[x_{i,j}\right]$} \\
\cline{3-8}
\multicolumn{2}{c|}{}  & $2.5\%$ & $5\%$ & $10\%$ & $50\%$ & $80\%$ &$100\%$ \\
\hline
\multirow{2}{*}{$I_1$} & Poisson & $0.025$ & $0.05$ & $0.10$ & $0.50$ & $0.80$ & $1.00$ \\
\cline{2-8}
 & Bernoulli & $0.025$ & $0.05$ & $0.09$ & $0.36$ & $0.48$ & $0.54$ \\
\hline
\multirow{2}{*}{$I_2$} & Poisson & $0.025$ & $0.05$ & $0.10$ & $0.50$ & $0.80$ & $1.00$ \\
\cline{2-8}
 & Bernoulli & $0.025$ & $0.05$ & $0.09$ & $0.38$ & $0.52$ & $0.59$ \\
\hline
\end{tabular}
\end{center}
\end{footnotesize}
\caption{Average number of detected photons per pixel $\textrm{E}\left[y_{i,j}\right]$ for the two images $I_1$ and $I_2$ corrupted by Poisson and Bernoulli noise for different values of $\textrm{E}\left[x_{i,j}\right]$.\label{tab:data_analysis}}
\vspace{-0.3cm}
\end{table}

\begin{table*}[!t]
\renewcommand{\arraystretch}{1.2}
\begin{footnotesize}
\begin{center}
\begin{tabular}{|c|c|c|c|c|c|c|c|}
\cline{3-8}                            
\multicolumn{2}{c|}{} &  \multicolumn{6}{|c|}{$\textrm{E}\left[x_{i,j}\right]$} \\
\cline{3-8}
\multicolumn{2}{c|}{}  & $2.5\%$ & $5\%$ & $10\%$ & $50\%$ & $80\%$ &$100\%$ \\
\hline                         
\multirow{4}{*}{$I_1$} & Ber-TV & $0.383$ $(0.533)$ & $0.309$ $(0.350)$ & $0.255$ $(0.214)$ & $\textbf{0.090}$ $(0.220)$ & $\textbf{0.068}$ $(0.174)$ & $\textbf{0.059}$ $(0.157)$ \\
\cline{2-8}
 & BID-GMRF & $\textbf{0.237}$ $(0.215)$ & $0.208$ $(0.184)$ & $0.179$ $(0.164)$ & $0.115$ $(0.125)$ & $0.091$ $(0.107)$ & $0.097$ $(0.115)$ \\
\cline{2-8}
 & NL-PCA & $0.286$ $(0.888)$ & $0.234$ $(0.207)$ & $\textbf{0.159}$ $(0.172)$ & $0.143$ $(0.135)$ & $0.206$ $(0.169)$ & $0.331$ $(0.200)$ \\
\cline{2-8}
 & PID-GMRF & $0.242$ $(0.227)$& $\textbf{0.207}$ $(0.180)$& $0.179$ $(0.149)$& $0.204$ $(0.143)$& $0.272$ $( 0.182)$& $0.320$ $(0.214)$\\
\hline
\hline                                                
\multirow{4}{*}{$I_2$} & Ber-TV & $0.191$ $(0.307)$ & $0.091$ $(0.157)$ & $0.065$ $(0.126)$ & $\textbf{0.042}$ $(0.088)$ & $\textbf{0.032}$ $(0.077)$ & $\textbf{0.032}$ $(0.077)$ \\
\cline{2-8}
 & BID-GMRF & $\textbf{0.081}$ $( 0.155)$ & $\textbf{0.071}$ $(0.142)$ & $0.066$ $(0.143)$ & $0.056$ $(0.103)$ & $0.037$ $(0.068)$ & $0.036$ $(0.071)$\\
\cline{2-8}
 & NL-PCA & $0.218$ $(0.757)$ & $0.169$ $(0.145)$ & $0.097$ $(0.092)$ & $0.049$ $(0.145)$ & $0.073$ $(0.228)$& $0.177$ $(0.283)$\\
\cline{2-8}
 & PID-GMRF & $0.095$ $(0.191)$ & $0.077$ $(0.150)$ & $\textbf{0.060}$ $(0.110)$ & $0.102$ $(0.156)$ & $0.173$ $(0.235)$ & $0.221$ $(0.287)$\\
\hline
\end{tabular}
\end{center}
\end{footnotesize}
\caption{Average normalized mean square errors (NMSEs) obtained by different algorithms for the images $I_1$ and $I_2$ corrupted by Bernoulli noise versus $\textrm{E}\left[y_{i,j}\right]$.\label{tab:MSE_Ber}}
\vspace{-0.3cm}
\end{table*}

We have compared our methods with the following state-of-the art methods: First, we considered a set of methods relying on the Poisson noise assumption. Precisely, we used SPIRAL-TV \cite{Harmany2012}, which solves the same optimization as PIDAL \cite{Figueiredo2010}; that is 
\begin{eqnarray}
\label{eq:poiss_TV}
\underset{\bfX \succeq \boldsymbol{0}}{\textrm{min}} -\log\left(f_0(\MATpix|\bfN,\bfX) \right) + \lambda_{\textrm{TV}} \sum_{t=1}^T\textrm{TV}\left(\bfX_t\right),
\end{eqnarray} 
where $\textrm{TV}\left(\cdot\right)$ is the total variation regularization whose influence is controlled by $\lambda_{\textrm{TV}}\geq 0$. We also applied the other regularizations proposed in \cite{Harmany2012} but SPIRAL-TV seems to provide the best and most robust results in this very sparse photon regime, which is why we only present the results obtained with this version of SPIRAL. We also implemented an alternative algorithm which solves the following problem 
\begin{eqnarray}
\label{eq:poiss_Lap}
\underset{\bfX \succeq \boldsymbol{0}}{\textrm{min}} -\log\left(f_0(\MATpix|\bfN,\bfX) \right) + \lambda_{\textrm{Lap}} \sum_{t=1}^T\norm{\bfD \bfx_t}_{2}^2,
\end{eqnarray} 
where $\bfx_t$ is the vectorized version of $\bfX_t$ and $\bfD$ is the $N_{\textrm{row}}\times N_{\textrm{col}}\times N_{\textrm{row}}N_{\textrm{col}}$ circulant convolution matrix of the Laplacian
filter \cite{Orieux2012}. In contrast to the TV regularization which promotes piece-wise constant intensity profiles, the penalization in \eqref{eq:poiss_Lap} promotes smooth intensity variations. 
The problem \eqref{eq:poiss_Lap} is solved using an ADMM scheme, similar to that used in PIDAL, therefore, this method is referred to as ``PIDAL-Lap''. We also used the NL-PCA algorithm \cite{Salmon2014} which is a state-of-the-art unsupervised method for image denoising under a Poisson noise assumption (we used the parameter values recommended in \cite{Salmon2014} in all the experiments presented here).

To the best of our knowledge, there is no published algorithm to directly estimate the intensity profiles involved in \eqref{eq:model_bern_joint}. However, we implemented an ADMM-based algorithm based on the following convex problem
\begin{eqnarray}
\label{eq:bern_TV}
\underset{\bfX \succeq \boldsymbol{0}}{\textrm{min}} -\log\left(f_1(\MATpix|\bfN,\bfX) \right) + \lambda_{\textrm{TV}} \sum_{t=1}^T\textrm{TV}\left(\bfX_t\right),
\end{eqnarray} 
This algorithm is referred to as ``Ber-TV'' in the remainder of this paper. Note that as described in \cite{Oh2015}, it might be possible possible to consider other regularizations, for both the Poisson and Bernoulli observation models. However, an extensive comparison of regularizations, potentially using changes of variables and whose regularization parameters need to be carefully adjusted, is outwith the scope of this paper and we concentrate on the TV and Laplacian-based penalizations, whose effects are easily understood and which require the adjustment of a single parameter.

We measure the performance of the different algorithms in term of normalized mean squared error (NMSE) defined by 
\begin{eqnarray}
\textrm{NMSE}_{t}=\dfrac{\sum_{i,j}(x_{i,j,t}-\hat{x}_{i,j,t})^2}{\sum_{i,j} x_{i,j,t}^2},
\end{eqnarray}
where $x_{i,j,t}$ (resp. $\hat{x}_{i,j,t}$) is the actual (resp. estimated) intensity value of the pixel $(i,j,t)$. The lower the NMSEs, the more similar the original and reconstructed images. Note that the NMSE does not depend on the intensity dynamic of the original image. We also evaluate how the reconstruction error varies across the image pixels around the NMSE using the standard deviation of the normalized squared error 
\begin{eqnarray}
\sqrt{\textrm{Var}\left[\dfrac{(x_{i,j,t}-\hat{x}_{i,j,t})^2}{\sum_{i,j} x_{i,j,t}^2/N_{\textrm{row}}N_{\textrm{col}}} \right]}.
\end{eqnarray}
  
We have applied the proposed PID-GMRF and BID-GMRF algorithms with $N_{\textrm{MC}}=2000$ (including $N_{\textrm{bi}}=600$ burn-in iterations) to the data corrupted by Poisson and Bernoulli noise. We have also applied PID-GMRF to data corrupted by Bernoulli noise to simulate the performance of methods relying on a Poisson noise assumption when denoising data recorded by non photon-number resolving detectors. The PIDAL-TV, PIDAL-Lap and Ber-TV algorithms, requires tuning of a regularization parameter ($\lambda_{\textrm{TV}}$ or $\lambda_{\textrm{Lap}}$). We adopted the methods proposed in \cite{Pereyra2015eusipco} to automatically adjust these hyperparameters, but these methods tend to significantly overestimate the smoothness of the intensity field, due to the extreme sparsity of the observed images and thus yield poor results, visually and in terms of NMSEs.
Consequently, for the results presented here, these hyperparameters have been optimized in a supervised manner in order to minimize the NMSE, which however requires knowing the actual intensity image in advance. 

\begin{figure}[h!]
\begin{minipage}[b]{.49\linewidth}
  \centering
\includegraphics[width=\linewidth]{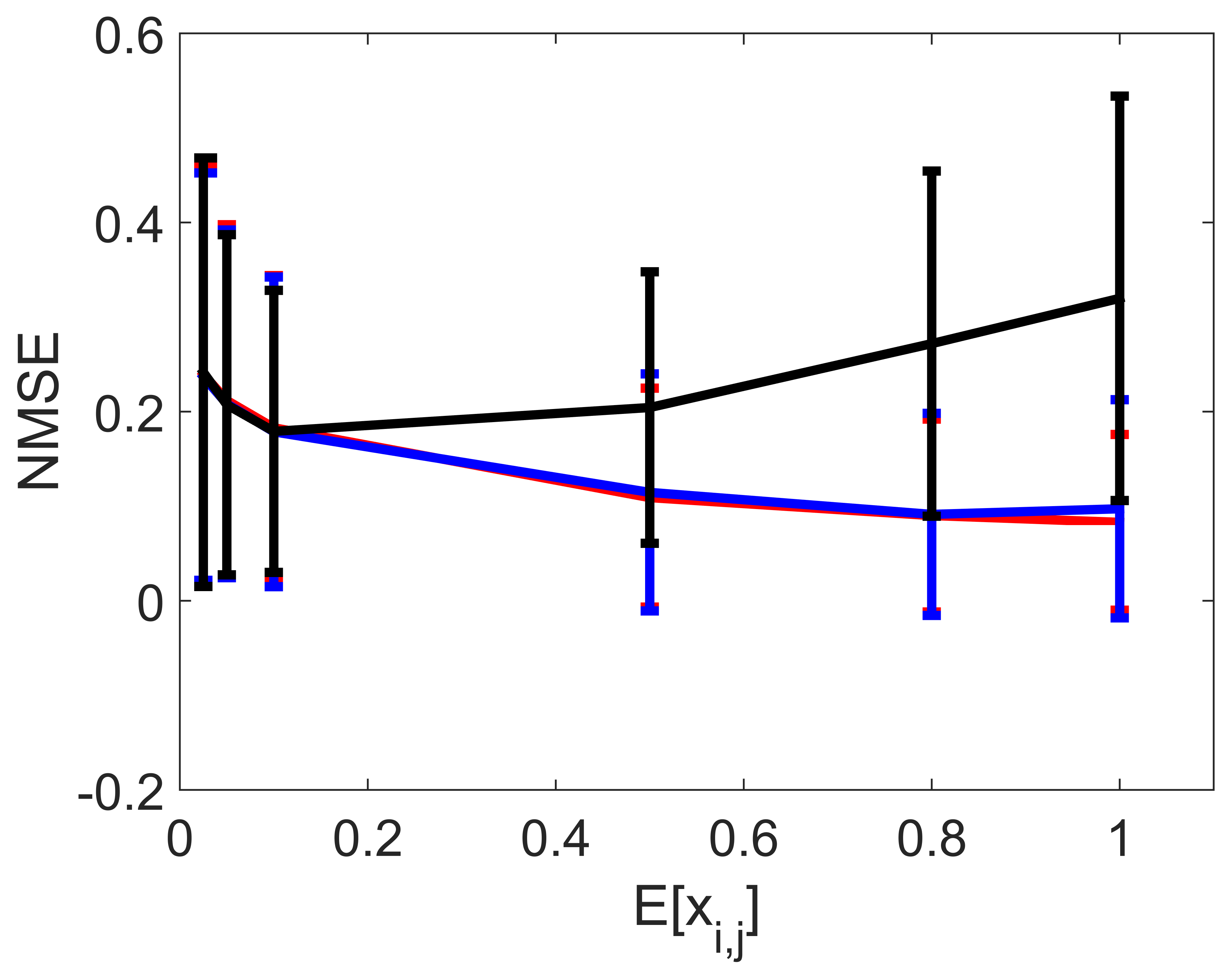}
 \vspace{-0.2cm}
  \centerline{(a) }\medskip
\end{minipage}
\hfill
\begin{minipage}[b]{0.49\linewidth}
  \centering
\includegraphics[width=\linewidth]{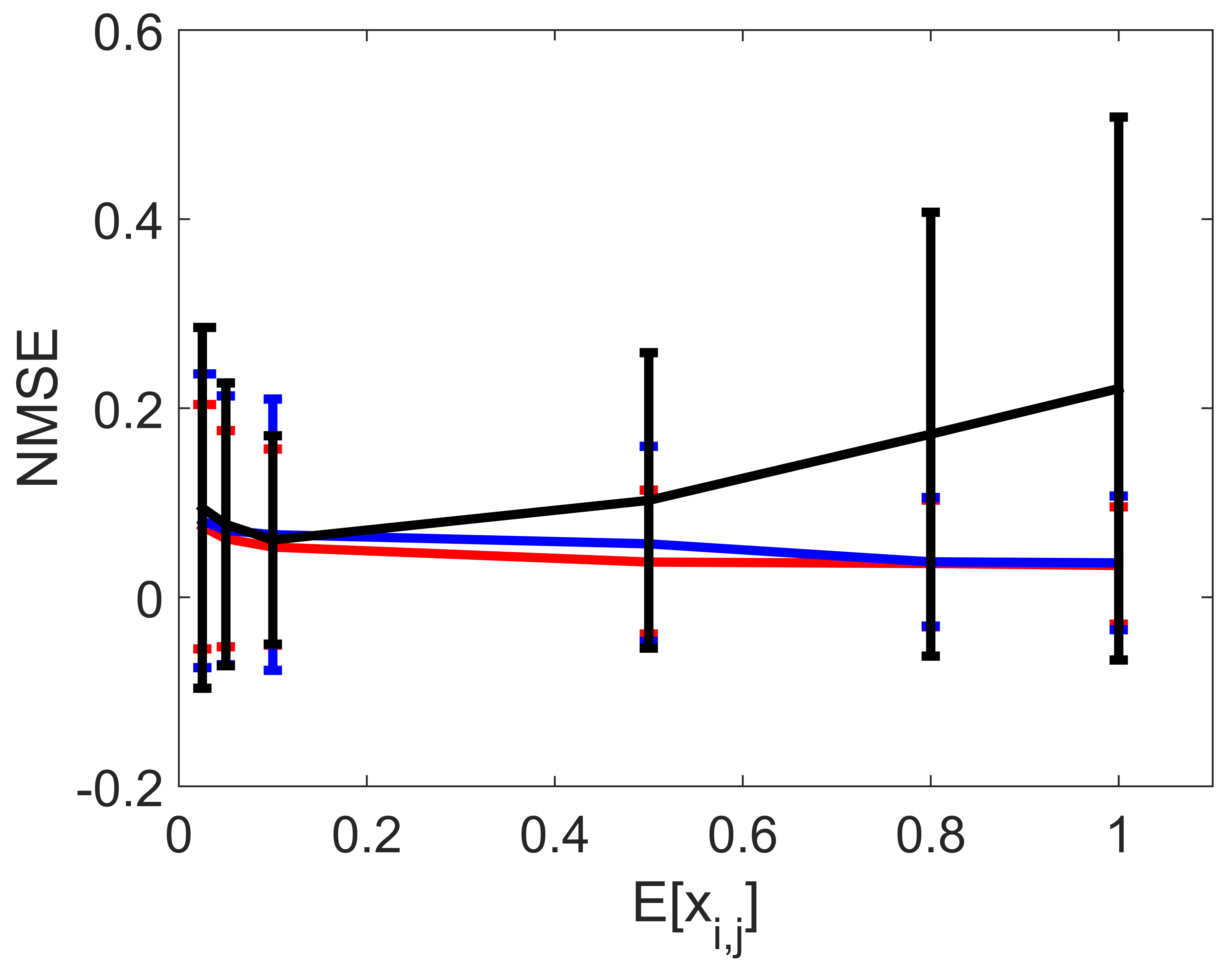}
 \vspace{-0.2cm}
  \centerline{(b)}\medskip
\end{minipage}
\vspace{-0.6cm} \caption{Average NMSEs obtained with PID-GMRF on images corrupted by Poisson noise (red lines), with BID-GMRF on images corrupted by Bernoulli noise (blue lines) and PID-GMRF on images corrupted by Bernoulli noise (black lines). The subplot (a) (resp. (b)) corresponds to $I_1$ (resp. $I_2$).} \label{fig:perf_single_im_MSE}
\end{figure}


\begin{table*}[ht!]
\renewcommand{\arraystretch}{1.2}
\begin{footnotesize}
\begin{center}
\begin{tabular}{|c|c|c|c|c|c|c|c|}
\cline{3-8}
\multicolumn{2}{c|}{} &  \multicolumn{6}{|c|}{$\textrm{E}\left[x_{i,j}\right]$} \\
\cline{3-8}
\multicolumn{2}{c|}{}  & $2.5\%$ & $5\%$ & $10\%$ & $50\%$ & $80\%$ &$100\%$ \\
\hline     
\multirow{4}{*}{$I_1$} & SPIRAL-TV & $0.410$ $(0.564)$ & $0.301$ $(0.366)$& $0.243$ $(0.360)$& $0.085$ $(0.221)$& $0.062$ $(0.176)$& $\textbf{0.051}$ $(0.143)$\\
\cline{2-8}
 & PIDAL-Lap & $0.258$ $(0.289)$& $0.221$ $(0.205)$& $0.150$ $(0.209)$& $\textbf{0.073}$ $(0.124)$& $\textbf{0.060}$ $(0.105)$& $0.054$ $(0.098)$\\
\cline{2-8}
 & NL-PCA & $0.346$ $(1.208)$& $\textbf{0.210}$ $(0.238)$& $\textbf{0.148}$ $(0.174)$& $0.077$ $(0.115)$& $0.073$ $(0.110)$& $0.072$ $(0.108)$\\
\cline{2-8}
 & PID-GMRF & $\textbf{0.240}$ $(0.220)$& $0.212$ $(0.186)$& $0.183$ $(0.161)$& $0.109$ $(0.116)$& $0.089$ $(0.102)$& $0.083$ $(0.093)$\\
\hline
\hline       
\multirow{4}{*}{$I_2$} & SPIRAL-TV & $0.167$ $(0.258)$& $0.083$ $(0.144)$& $0.067$ $(0.128)$& $0.028$ $(0.071)$& $0.027$ $(0.070)$& $0.025$ $(0.061)$\\
\cline{2-8}
 & PIDAL-Lap & $\textbf{0.065}$ $(0.124)$& $\textbf{0.052}$ $(0.104)$& $\textbf{0.042}$ $(0.088)$& $0.026$ $(0.064)$& $0.022$ $(0.057)$& $0.021$ $(0.052)$\\
\cline{2-8}
 & NL-PCA & $0.186$ $(0.938)$& $0.075$ $(0.174)$& $0.048$ $(0.092)$& $\textbf{0.022}$ $(0.054)$& $\textbf{0.019}$ $(0.050)$& $\textbf{0.018}$ $(0.049)$\\
\cline{2-8}
 & PID-GMRF & $0.075$ $(0.129)$& $0.062$ $(0.114)$& $0.053$ $(0.104)$& $0.037$ $(0.076)$& $0.036$ $(0.067)$& $0.034$ $(0.062)$\\
\hline
\end{tabular}
\end{center}
\end{footnotesize}
\caption{Average normalized mean square errors (NMSEs) obtained by different algorithms for the images $I_1$ and $I_2$ corrupted by Poisson noise versus $\textrm{E}\left[y_{i,j}\right]$.\label{tab:MSE_Poiss}}
\vspace{-0.3cm}
\end{table*}

We first compare the performance of the two proposed methods when denoising images corrupted by Bernoulli and Poisson noise. Fig. \ref{fig:perf_single_im_MSE} depicts the NMSEs and associated confidence regions ($\pm$ standard deviation), computed for each image and averaged over the $T=20$ noise realizations. This figure shows that when $\textrm{E}\left[x_{i,j}\right]\rightarrow 0$ the results obtained by the two algorithms are similar for the two observation models. This can be explained by the fact that the likelihoods \eqref{eq:model_poisson} and \eqref{eq:lik_bern} become similar and weakly informative and that the intensity estimates are mostly driven by the intensity prior model, which is the same in all the scenarios. As expected, using a Poisson observation model when the data are Bernoulli distributed (black lines) yields less accurate intensity estimates when $\textrm{E}\left[x_{i,j}\right]$ increases due to the poor Poisson approximation of the Bernoulli distribution. More surprisingly, the two methods using the correct observation model (blue and red curves) present similar behaviors for all values of $\textrm{E}\left[x_{i,j}\right]$. Indeed, we could expect the intensity estimation to be significantly less accurate when considering Bernoulli observations since the detectors cannot detect all the photons reaching the sensors (at most one per pixel). However, these results show that although a non photon-number resolving detector is used, it is possible to obtain similar intensity estimates to those obtained by an ideal detector (provided that the appropriate observation model is used and that $\textrm{E}\left[x_{i,j}\right]$ is not too large).

Table \ref{tab:MSE_Ber} compare the NMSEs and associated standard deviations (averaged over $T=20$ realizations) obtained by Ber-TV, BID-GMRF, NL-PCA and PID-GMRF when denoising the images $I_1$ and $I_2$ corrupted by Bernoulli noise. These results confirm that for high values of $\textrm{E}\left[x_{i,j}\right]$, the methods relying on Poisson noise assumption (NL-PCA and PID-GMRF) provide less accurate intensity estimates than Ber-TV and BID-GMRF. Moreover, BID-GMRF is more robust than Ber-TV for small values of $\textrm{E}\left[x_{i,j}\right]$ but is outperformed by Ber-TV in terms of NMSE (when appropriately tuned) when $\textrm{E}\left[x_{i,j}\right] \rightarrow 1$. It is important to mention that the NMSE performance has to be moderated by the relatively high standard deviations in Table \ref{tab:MSE_Ber}. This table also shows that even if BID-GMRF does not necessarily provide lower NMSEs, it generally provides lower standard deviations, in particular for small values of $\textrm{E}\left[x_{i,j}\right]$.

\begin{figure}[h!]
  \centering
  \includegraphics[width=\columnwidth]{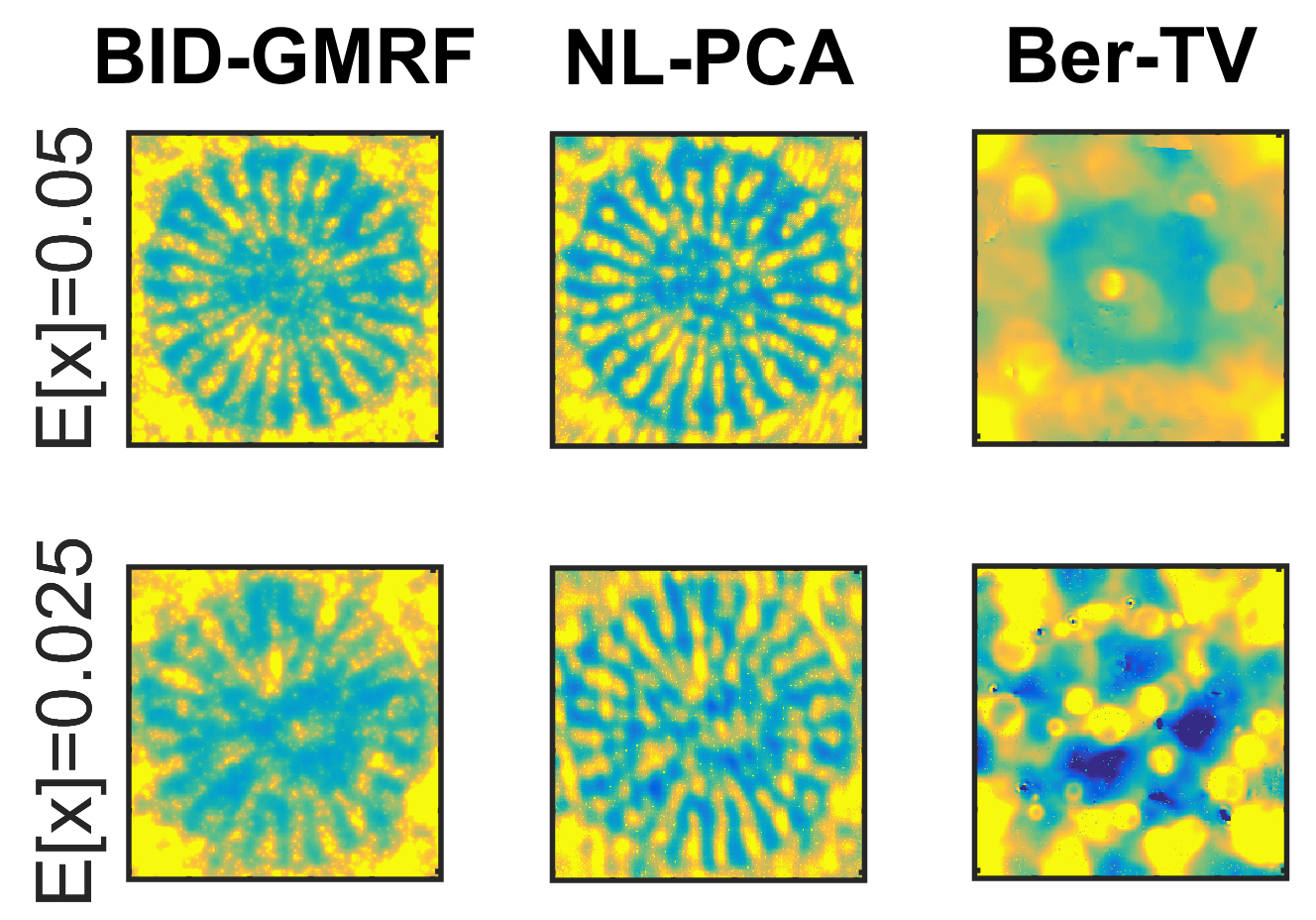}
	\vspace{-0.6cm}
  \caption{Intensity estimates for $I_1$ (corrupted using \eqref{eq:lik_bern}), using BID-GMRF, NL-PCA and Ber-TV, and for different values of $\textrm{E}\left[x_{i,j}\right]$. For each row, the same scale $(0,\textrm{max}(x_{i,j}))$ is used for the three methods.}
  \label{fig:perf_pattern_3algos}
\end{figure}

\begin{figure}[h!]
  \centering
  \includegraphics[width=\columnwidth]{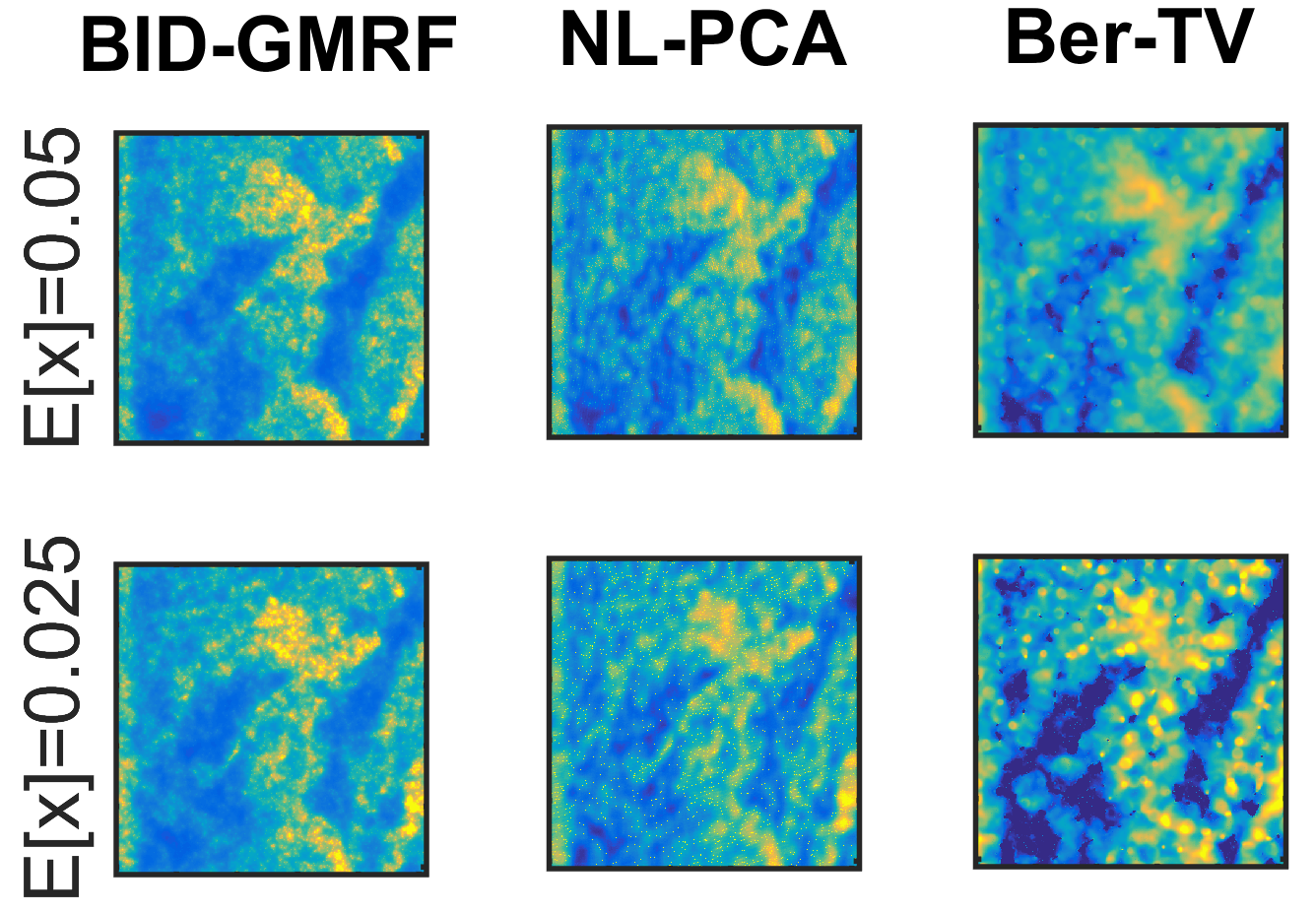}
	\vspace{-0.6cm}
  \caption{Intensity estimates for $I_2$ (corrupted using \eqref{eq:lik_bern}), using BID-GMRF, NL-PCA and Ber-TV, and for different values of $\textrm{E}\left[x_{i,j}\right]$. For each row, the same scale $(0,\textrm{max}(x_{i,j}))$ is used for the three methods.}
  \label{fig:perf_lena_3algos}
\end{figure}

Figs. \ref{fig:perf_pattern_3algos} and \ref{fig:perf_lena_3algos} compare examples of single image denoising using $\textrm{E}\left[x_{i,j}\right]=5\%$ and $\textrm{E}\left[x_{i,j}\right]=2.5\%$. These results illustrate the fact the GMRF considered is flexible enough to capture the spatial correlation of piece-wise constant ($I_1$) and smoother ($I_2$) images and is visually more robust than Ber-TV (less prominent patch-like artifacts). NL-PCA provides similar images and is able to detect spatial structure in the data but underestimate the large intensities due to the model mismatch, yielding higher NMSE when $\textrm{E}\left[x_{i,j}\right] \rightarrow 1$ (see Table \ref{tab:MSE_Ber}). 

Table \ref{tab:MSE_Poiss} compares the NMSEs (averaged over $T=20$ realizations) obtained by SPIRAL-TV, PIDAL-Lap, NL-PCA and PID-GMRF when denoising the images $I_1$ and $I_2$ corrupted by Poisson noise. These results shows that the proposed Bayesian approach, when assuming Poisson noise, provides more robust results than the other state-of-the-art methods when $\textrm{E}\left[y_{i,j}\right] \rightarrow 0$. When $\textrm{E}\left[y_{i,j}\right] \rightarrow 1$ however, the three other methods generally yield slightly better NMSEs. Although SPIRAL-TV and PIDAL-Lap need to be tuned to obtain such performance, NL-PCA does not which open routes to further improve the denoising performance of BID-GMRF, e.g., using dictionary techniques such as NL-PCA, in particular when $\textrm{E}\left[y_{i,j}\right] \rightarrow 1$.

\subsection{Denoising of image sequences}

\begin{figure*}[!t]
\begin{minipage}[b]{.32\linewidth}
  \centering
\includegraphics[width=5.5cm]{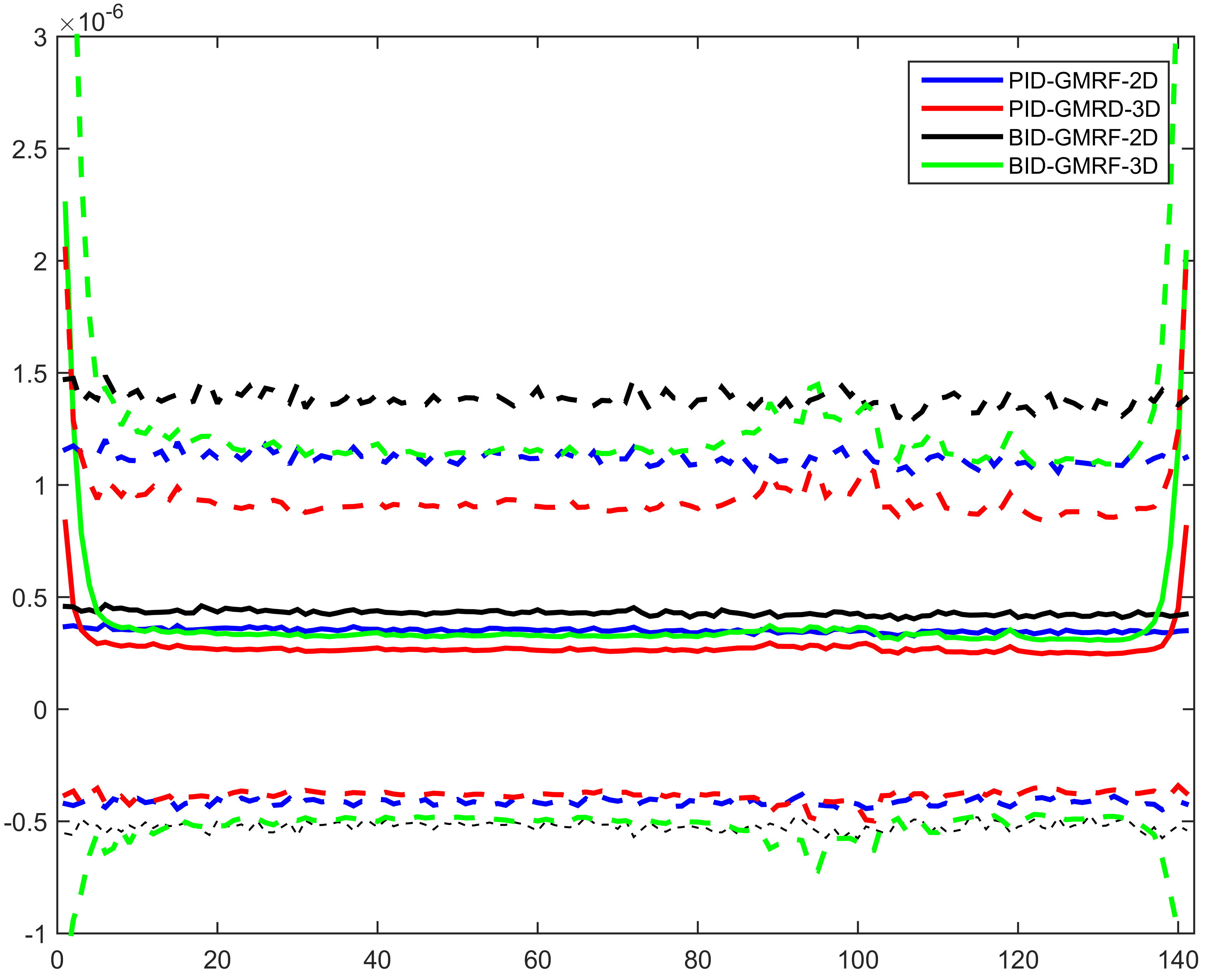}
 \vspace{-0.2cm}
  \centerline{(a) }\medskip
\end{minipage}
\hfill
\begin{minipage}[b]{0.32\linewidth}
  \centering
\includegraphics[width=5.5cm]{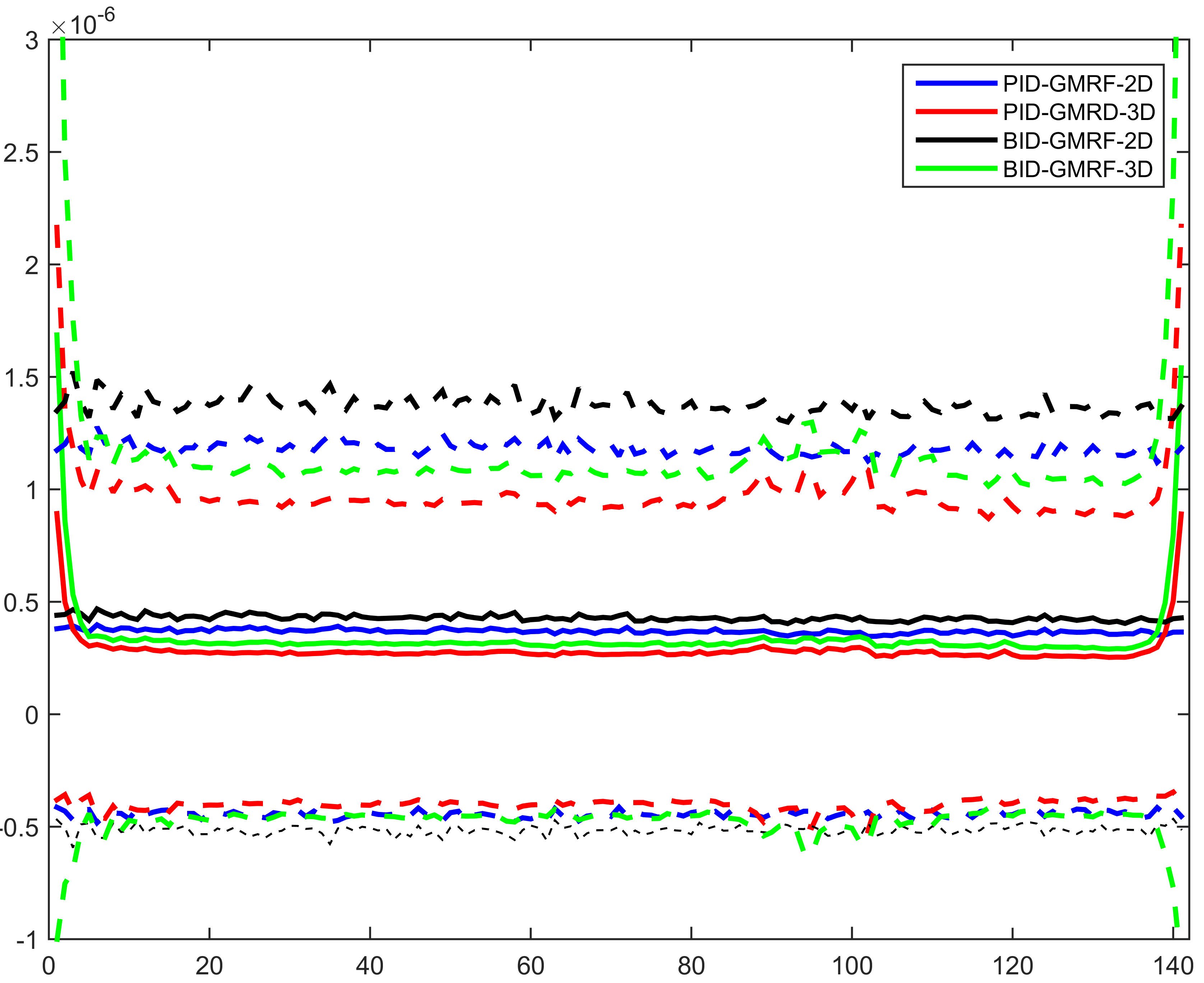}
 \vspace{-0.2cm}
  \centerline{(b)}\medskip
\end{minipage}
\hfill
\begin{minipage}[b]{.32\linewidth}
  \centering
\includegraphics[width=5.5cm]{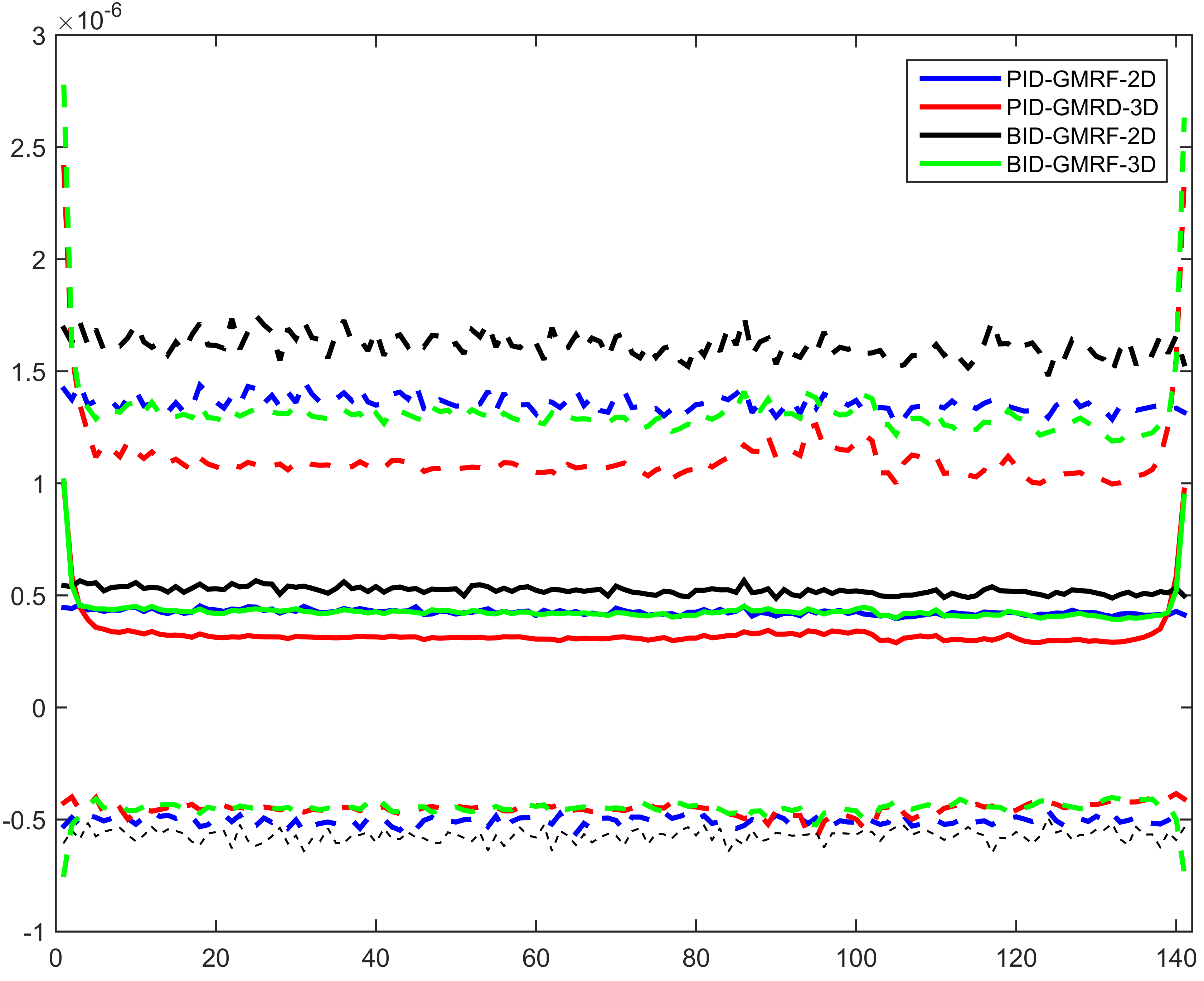}
 \vspace{-0.2cm}
  \centerline{(c) }\medskip
\end{minipage}

\begin{minipage}[b]{0.32\linewidth}
  \centering
\includegraphics[width=5.5cm]{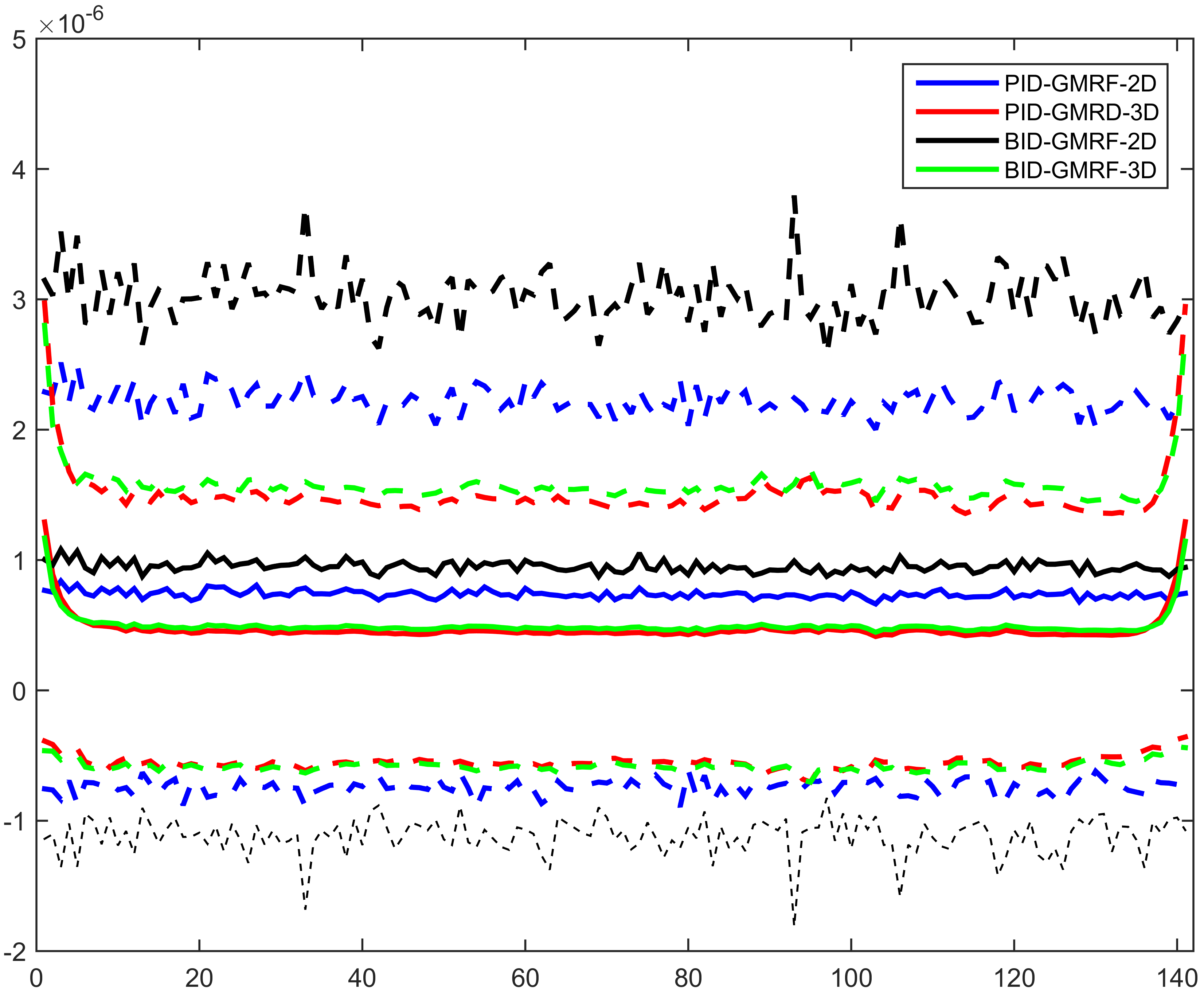}
 \vspace{-0.2cm}
  \centerline{(d)}\medskip
\end{minipage}
\hfill
\begin{minipage}[b]{.32\linewidth}
  \centering
\includegraphics[width=5.5cm]{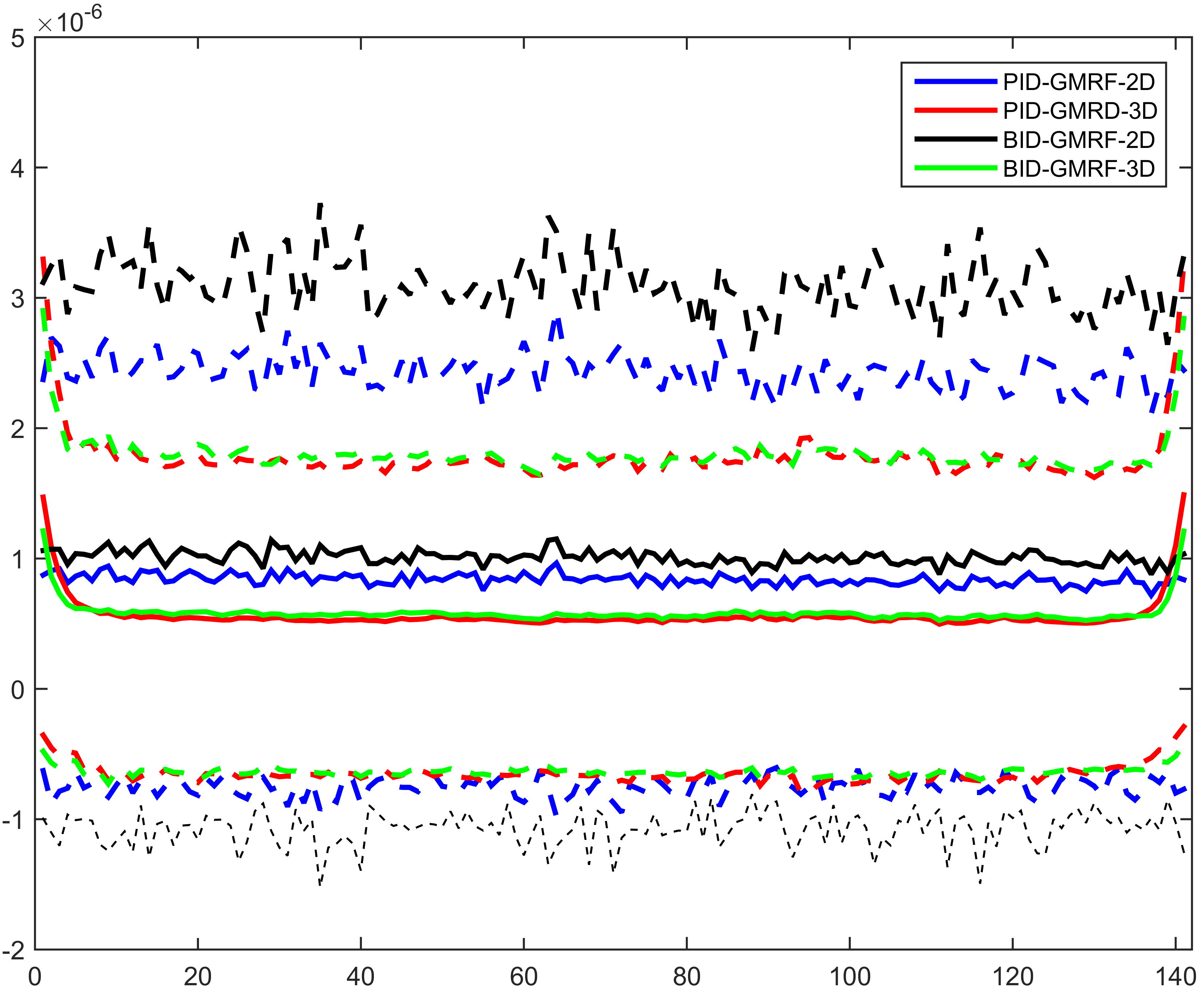}
 \vspace{-0.2cm}
  \centerline{(e) }\medskip
\end{minipage}
\hfill
\begin{minipage}[b]{0.32\linewidth}
  \centering
\includegraphics[width=5.5cm]{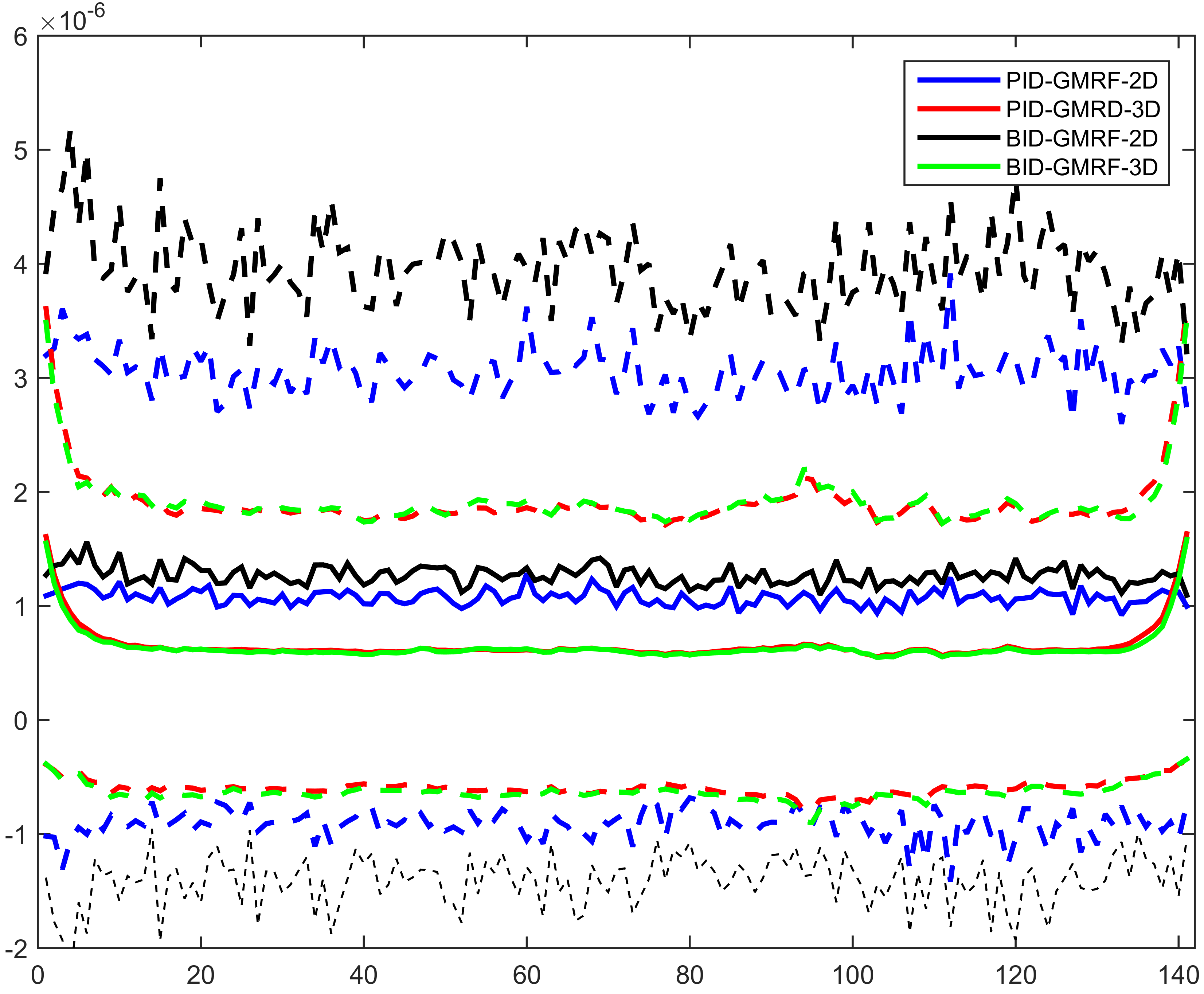}
 \vspace{-0.2cm}
  \centerline{(f)}\medskip
\end{minipage}
\caption{NMSEs obtained with PID-GMRF and BIP-GMRF (2D and 3D versions) on the synthetic videos of the xylophone for $\textrm{E}\left[x_{i,j}\right]=1$ (a), $\textrm{E}\left[x_{i,j}\right]=0.8$ (b), $\textrm{E}\left[x_{i,j}\right]=0.5$ (c), $\textrm{E}\left[x_{i,j}\right]=0.1$ (d), $\textrm{E}\left[x_{i,j}\right]=0.05$ (e) and $\textrm{E}\left[x_{i,j}\right]=0.025$ (f).} \label{fig:perf_video}
\end{figure*}

We now illustrate the benefits of the proposed 3D GMRF model when denoising videos constructed from single-photon data. We consider a video composed of $T=141$ frames of size $240 \times 310$ pixels, which represents someone striking a xylophone. This video has been selected from the video library available in Matlab R2014b. In the experiments presented in this section, there is no missing data and we used $\eta_{i,j}=1, \forall (i,j)$. The original video has been scaled such that the expected number of counts (averaged over the image pixels and frames) $\textrm{E}\left[x_{i,j,t}\right] \in \{2.5\%;5\%;10\%;50\%;80\%;100\%\}$. 
We have applied the proposed PID-GMRF (resp. BID-GMRF) algorithms with $N_{\textrm{MC}}=3000$, including $N_{\textrm{bi}}=1000$ burn-in iterations,  to the data corrupted by Poisson (resp. Bernoulli) noise. The methods using 2D (resp. 3D) GMRFs are denoted PID-GMRF-2D and BID-GMRF-2D (resp. PID-GMRF-3D and BID-GMRF-3D). Fig. \ref{fig:perf_video} compares the NMSEs, obtained by the proposed algorithms (using the correct observation model). These plots show that the NMSEs generally increase as $\textrm{E}\left[x_{i,j,t}\right]$ decreases and that the NMSEs are similar across the $T$ frames when using PID-GMRF-2D and BID-GMRF-2D. When PID-GMRF-3D and BID-GMRF-3D are used instead, the NMSEs generally decrease due to the consideration of the temporal correlation between successive frames. Note that the NMSEs increase at the very beginning and the very end of the sequences due to the GMRF boundary conditions considered. This bias can however be easily reduced if we further assume that the temporal sequence is cyclic. In order not to add unnecessary assumptions in the general case, we did not present this case (which can be addressed by changing the GMRF boundary conditions (see Fig. \ref{fig:neighbour_GMRF2})).

%
%

\section{Simulations using real data}
\label{sec:simus_real}
\begin{figure}[h!]
\centering
\includegraphics[width=\columnwidth]{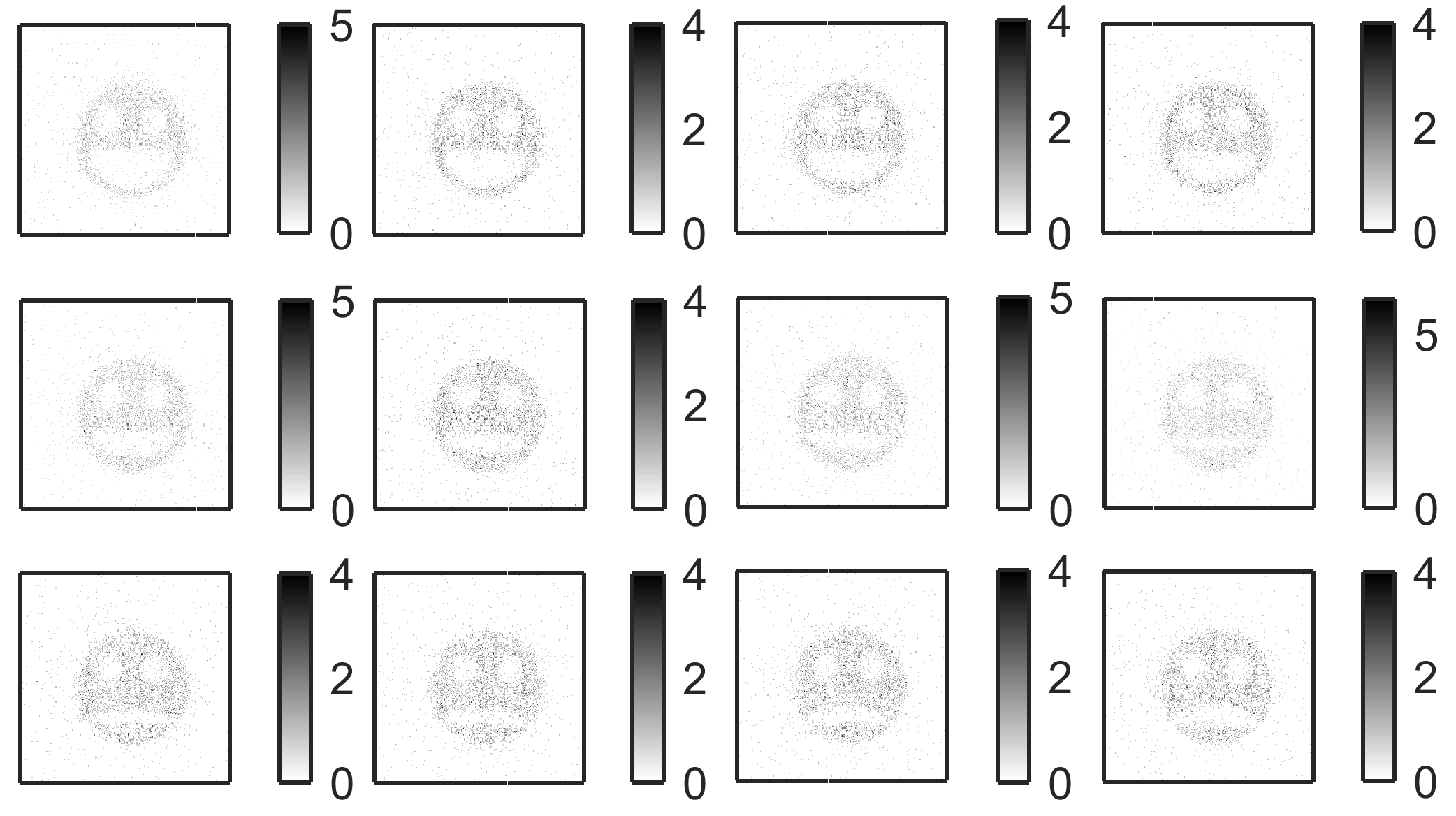}
\vspace{-0.6cm} \caption{Measured photon counts, obtained by integration of groups of $300$ successive images.} \label{fig:image_real}
\end{figure}
We illustrate the benefits of the proposed denoising framework to denoise sparse images of a dynamic object recorded by a ghost-imaging system similar to those considered in \cite{Tasca2013,Aspden2016}. The system considered here uses correlated photons at $710$nm and the images were displayed on a spatial light modulator (SLM). We consider a set of $12$ spatial patterns, i.e, $12$ smiley faces  gradually changing from a sad to happy face. The images of size $256 \times 256$ are recorded by an intensified camera with a CCD detector array (ICCD) triggered by a Perkin Elmer silicon SPAD  (see \cite{Tasca2013,Aspden2016} for more details about data acquisition and setup of the ghost-imaging instrument). Each face is observed over $300$ seconds with a frame rate of $1$Hz, leading to $300$ frames per face position. The ICCD acts here as a non photon-number resolving SPD, and thus provides binary images. The average intensity profile relates to the image of the faces formed from a polished silicon wafer onto which was patterned a microscopic gold test target. At the wavelength considered, the silicon is transparent whereas the gold
layer is not. Consequently, the acquired images are darker in the region where the gold target is present.  The laser source is adjusted so that the average number of detected photons per pixel and per frame is significantly lower than $5\%$ ($\textrm{E}\left[x_{i,j,t}\right]=2.3 \times 10^{-4}$ and $x_{i,j,t}<2\%, \forall (i,j,t)$). In other word, the probability of having more than one photon reaching a given pixel within a given $1s$ frame is extremely low. In this extremely sparse photon-limited imaging regime, the distributions of the photon count can be approximated by Poisson distributions. Fig. \ref{fig:image_real} depicts the accumulated photon counts obtained by summing the $300$ images associated with each position.

\begin{table}[ht!]
\renewcommand{\arraystretch}{1.2}
\begin{footnotesize}
\begin{center}
\begin{tabular}{|c|c|c|c|c|}
\cline{2-5}
\multicolumn{1}{c|}{} &  \multicolumn{4}{|c|}{Integration time per frame (in seconds)} \\
\cline{2-5}
\multicolumn{1}{c|}{}  & $25$ & $50$ & $100$ & $300$\\
\hline     
Per frame photon counts & $442.3$ & $844.7$ & $1689.4$& $5068.1$\\
\hline
Per frame detection counts & $417.3$ & $824.0$ & $1608.7$ & $4379.3$\\
\hline
Per pixel detection rate & $0.64\%$ & $1.26\%$ & $2.45\%$  & $6.68\%$\\
\hline
\hline
NMSE & $0.220$ & $0.071$  & $0.050$ & $0.017$\\
\hline
\end{tabular}
\end{center}
\end{footnotesize}
\caption{Average normalized mean square errors (NMSEs) and standard deviations (in brackets) ($\times 10^{-2}$) obtained for different exposure times/number of frames per group.\label{tab:MSE_real}}
\vspace{-0.3cm}
\end{table}

\begin{figure}[h!]
  \centering
  \includegraphics[width=\columnwidth]{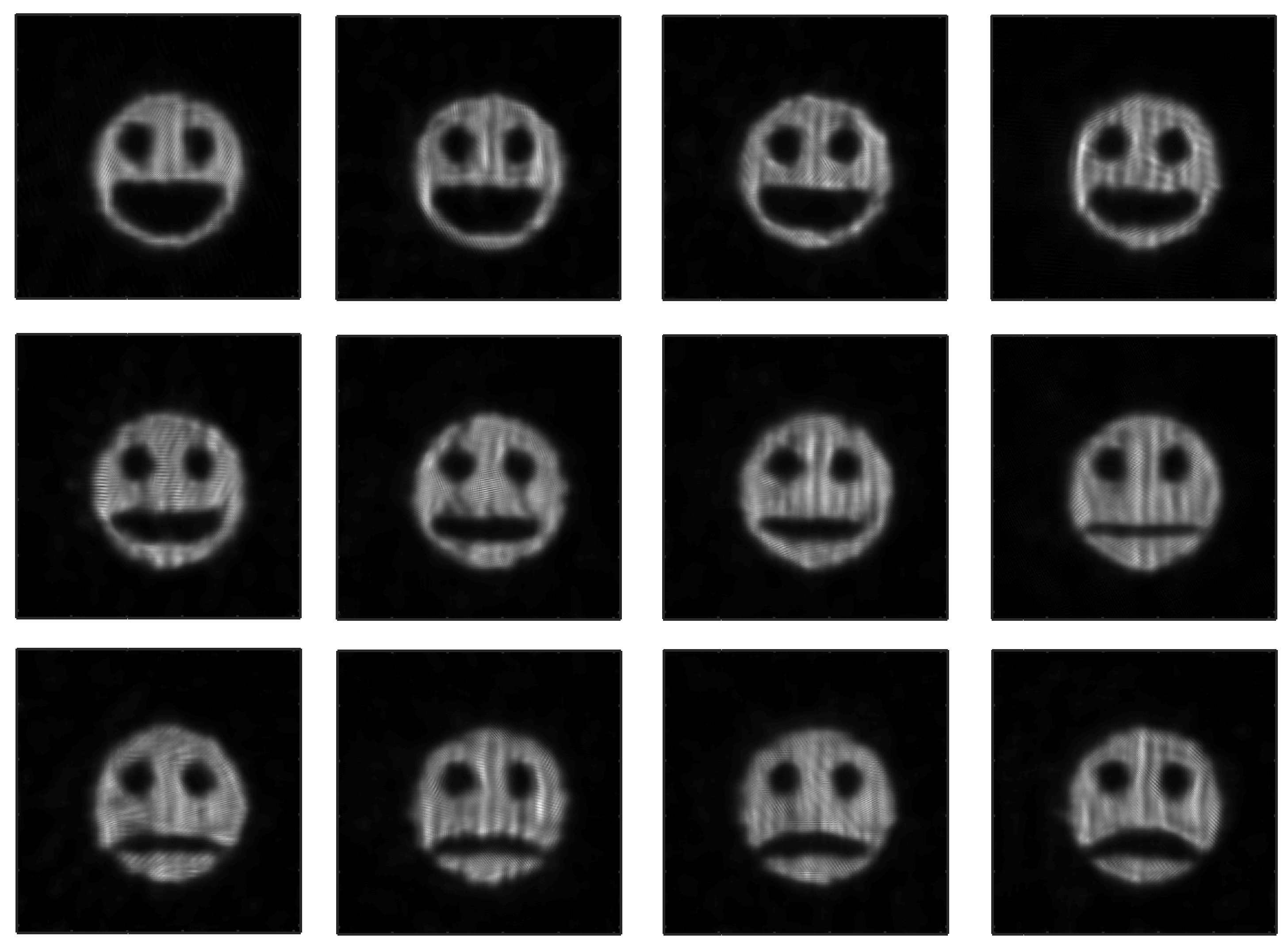}
	\vspace{-0.6cm}
  \caption{Images denoised using NL-PCA and used as reference to compute NMSEs. The input images are obtained by summing groups of $300$ successive original images during which the intensity field is stationary.}
  \label{fig:ref_real}
\end{figure}
\begin{figure}[h!]
  \centering
  \includegraphics[width=\columnwidth]{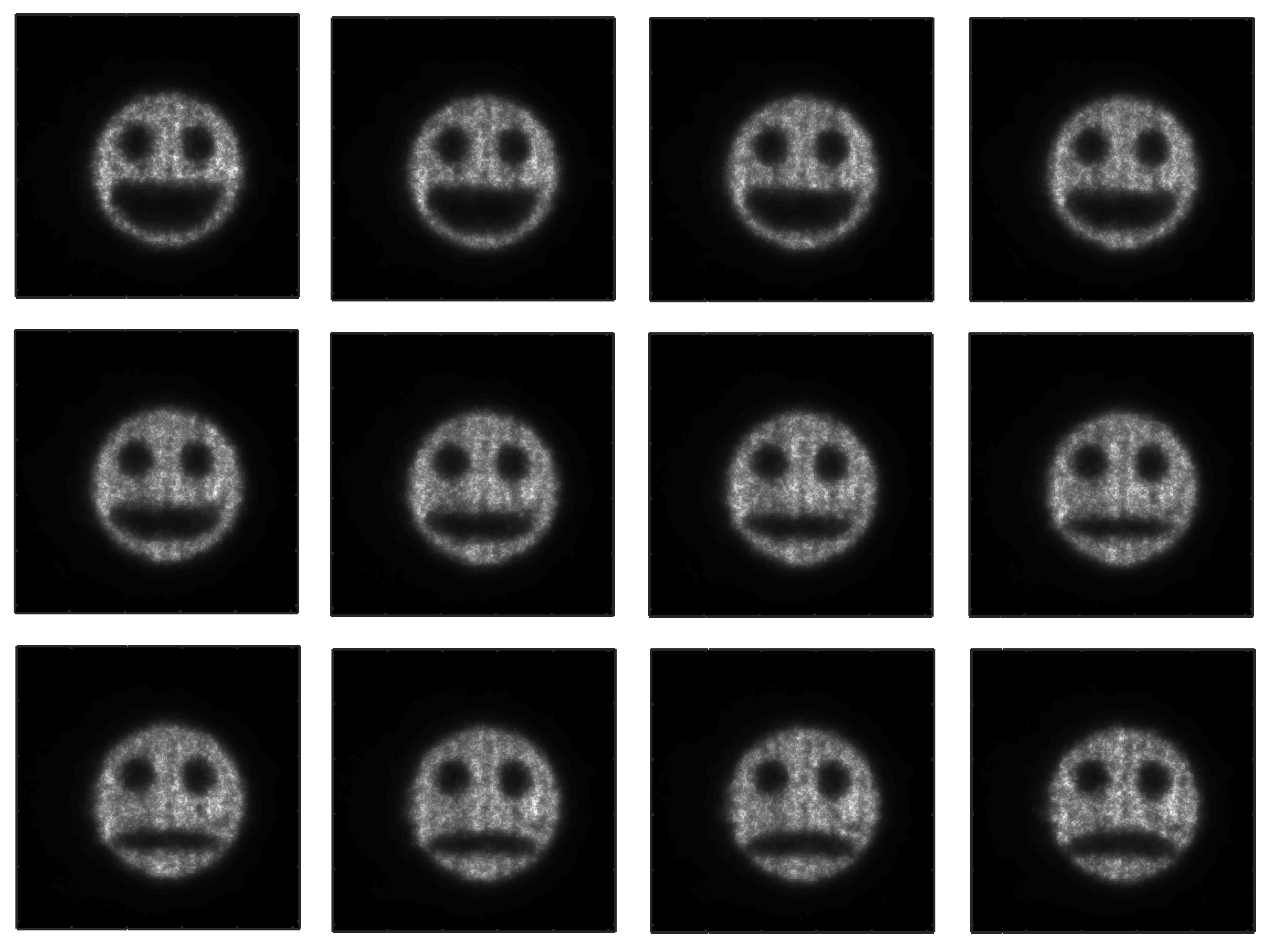}
	\vspace{-0.6cm}
  \caption{Examples of images using BID-GMRF-3D. The input images are obtained by summing groups of $300$ successive original images, during which the underlying intensity field is stationary. The images are then thresholded (presence/absence of detected photons) to simulate longer integration times ($300$s here).}
  \label{fig:example_real_300}
\end{figure}
\begin{figure}[h!]
  \centering
  \includegraphics[width=\columnwidth]{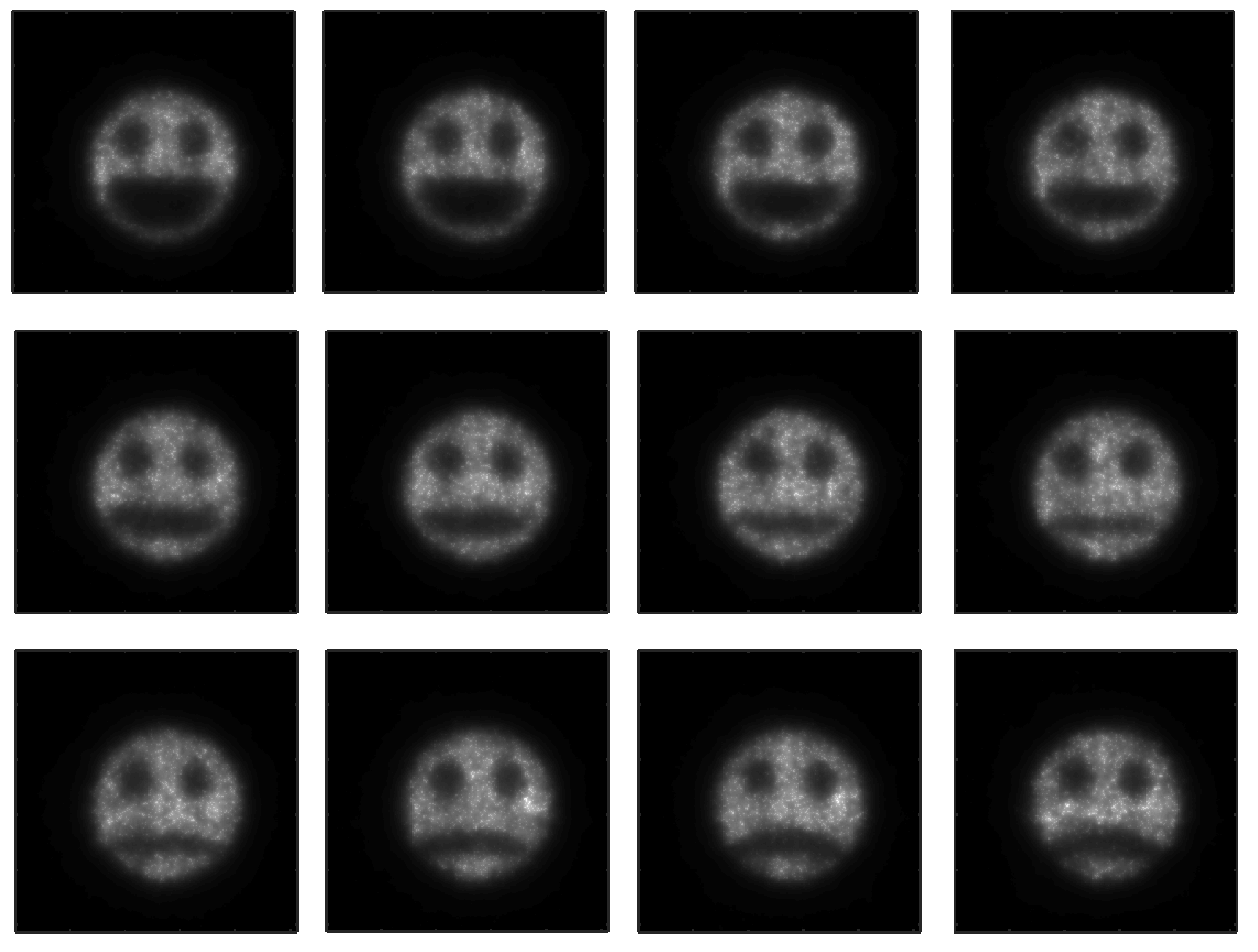}
	\vspace{-0.6cm}
  \caption{Examples of images using BID-GMRF-3D. The input images are obtained by summing groups of $50$ successive original images, during which the underlying intensity field is stationary. The images are then thresholded (presence/absence of detected photons) to simulate longer integration times ($50$s here). Each face position is thus visible in six successive images and the images presented correspond to the first image of each position.}
  \label{fig:example_real_50}
\end{figure}
\begin{figure}[h!]
  \centering
  \includegraphics[width=\columnwidth]{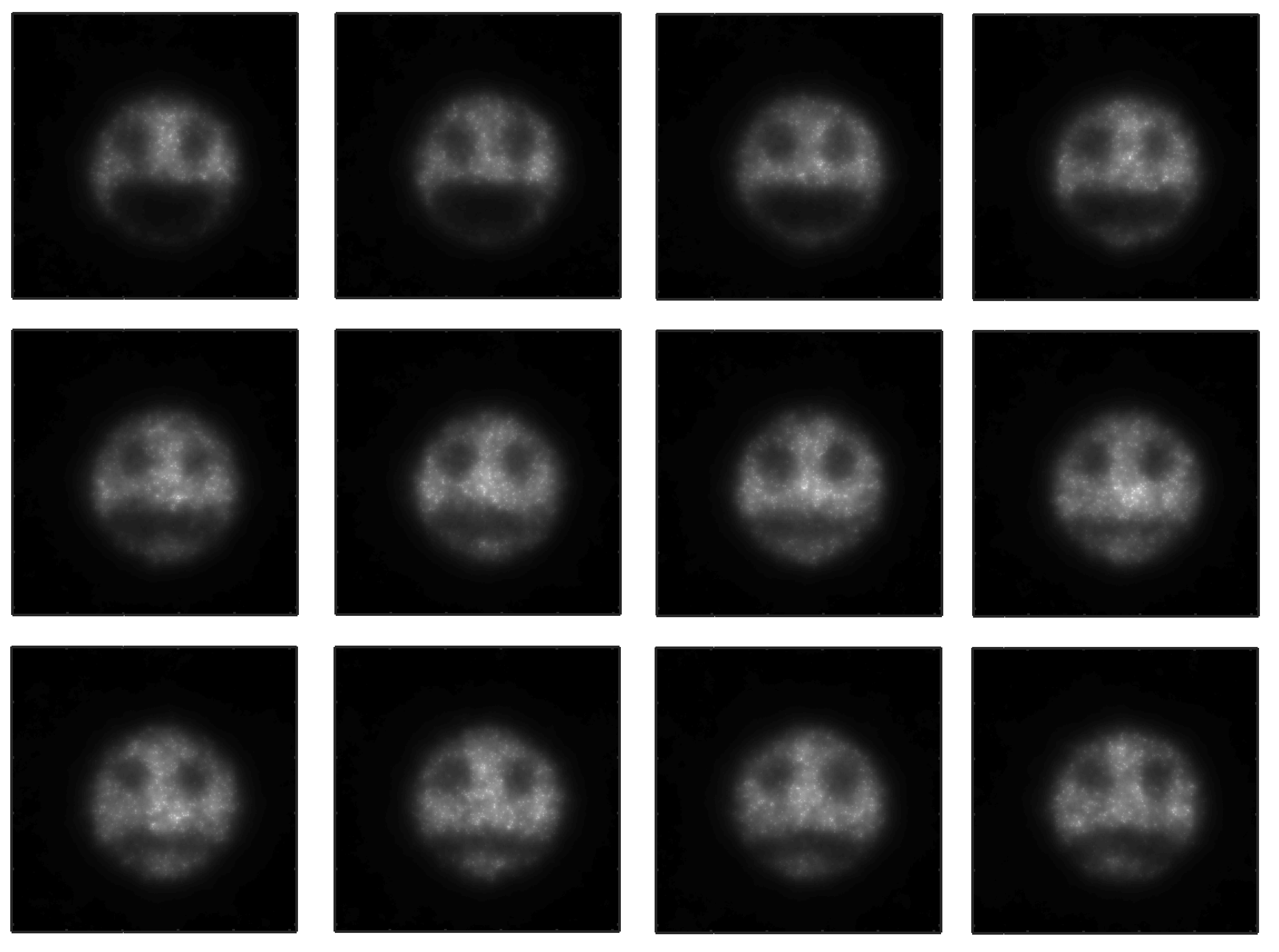}
	\vspace{-0.6cm}
  \caption{Examples of images using BID-GMRF-3D. The input images are obtained by summing groups of $25$ successive original images, during which the underlying intensity field is stationary. The images are then thresholded (presence/absence of detected photons) to simulate longer integration times ($25$s here). Each face position is thus visible in twelve successive images and the images presented correspond to the first image of each position.}
  \label{fig:example_real_25}
\end{figure}

To illustrate the benefits of the proposed denoising method, we denoise images that would have been obtained using exposure times of $25$s, $50$s, $100$s and $300$s. Such images are obtained by integrating non-overlapping groups of $25$ up to $300$ images. These images (approximately corrupted by Poisson noise) are then used to produce binary images associated with the presence/absence of detected photons within successive $25$s, $50$s, $100$s or $300$s periods. The top rows of Table \ref{tab:MSE_real} show the average number of photons in the images to be enhanced. In particular, less than $500$ photons per frame are available when considering the shortest exposure, which corresponds to a per pixel detection rate less than $1\%$. The bottom row of Table \ref{tab:MSE_real} compares the NMSEs obtained using BID-GMRF-3D as denoising method, where the reference intensities, depicted in Fig. \ref{fig:ref_real}, are those obtained with NL-PCA on the integrated groups of $300$ images. Note that the images depicted in Fig. \ref{fig:ref_real} present some vertical artifacts and should not be considered as absolute ground truth. However, since NL-PCA provides the visually most accurate enhanced images (over all the existing methods considered in Section \ref{sec:simus_synth}), this algorithm has been used as reference. For completeness, examples of denoised images extracted from the whole image sequences are provided in Figs. \ref{fig:example_real_300}, \ref{fig:example_real_50} and \ref{fig:example_real_25}. These results show that it is possible to use much lower frame rates (combined with lower overall exposures) and still obtain satisfactory intensity field estimates, which can be particularly useful to reduce the amount of data (divided by up to $300$ here) to be stored, transmitted and/or processed. These results also show that the proposed method can be used to enhance image sequences of dynamic scenes constructed from extremely sparse single-photon data.

\section{Conclusion}
\label{sec:conclusion}
Here we have proposed a new Bayesian method for binary image denoising. The model considered assumed that each pixel measurement follows a Bernoulli distribution whose mean is related by a nonlinear function to the underlying intensity value to be recovered. In contrast with classical Poisson noise models, this model is particularly adapted for data recorded single-photon detectors which are not photon-number resolving, especially when the unknown mean intensity value tends to $1$. A gamma Markov random field was proposed to design an intensity prior model able and capture the spatial and temporal structures of the unknown intensity field. A Markov chain Monte Carlo method was then developed to exploit the resulting posterior distribution and estimate the parameters of interest, including the regularization parameters of the Markov random field (thus avoiding parameter tuning via cross-validation). By including a minor modification of the algorithm, we have shown that the proposed method can also be applied to data corrupted by Poisson noise. A series of simulations conducted on synthetic data demonstrated the benefits (robustness) of the proposed method, especially for extremely sparse data. Moreover, we have demonstrated that the proposed version assuming Poisson noise is able to compete with state-of-the art denoising methods (based on a Poisson noise assumption). We have shown that for average intensities close to 1, it is possible to obtain from saturating sensors, an estimation accuracy close to that obtained using non-saturating sensors. For instance, the results of simulations conducted using real sparse single-photon measurements illustrated how one can reduce the amount of data (by reducing the frame rate here but one could also adjust the laser source and reduce the overall acquisition time when possible) while being able to estimate the intensity profile without a significant performance degradation.

Here we used a hidden gamma Markov random field to build a prior model. However, we noticed that dictionary learning techniques (such as NL-PCA) can significantly improve the denoising performance in the presence of sparse single-photon images. Including such considerations in 
future binary image denoising methods is clearly interesting. Moreover, the generalization of the proposed methodology for images following binomial distributions (e.g., sum of binary images) is currently under investigation.  
\newpage
\bibliographystyle{IEEEtran}
\bibliography{biblio}

\end{document}

%% file: notations_yoann.tex
\input com_notations.tex

\newcommand{\MATpix}{\mathbf{Y}}
\newcommand{\pix}[2]{y_{#1,#2}}

\newcommand{\norm}[1]{\left\|#1\right\|}

\newenvironment{algogo}[1]{
\smallskip
\noindent \hrule\vspace{0.2\baselineskip} \hrule
\begin{small}
\refstepcounter{algo} \center{\bf \textsc{Algorithm \thealgo}}
\\{\center{\bf #1}}
\smallskip
\flushleft
 } {
\end{small}
\smallskip
\hrule\vspace{0.2\baselineskip} \hrule
\smallskip
}

\newcounter{algo}
\renewcommand{\thealgo}{\arabic{algo}}

%% file: com_notations.tex
\def\bfx{{\mathbf{x}}}

\def\bfD{{\mathbf{D}}}

\def\bfH{{\mathbf{H}}}

\def\bfN{{\mathbf{N}}}

\def\bfU{{\mathbf{U}}}

\def\bfW{{\mathbf{W}}}
\def\bfX{{\mathbf{X}}}

